\documentclass[aps,prl,twocolumn,amsmath,amssymb,floatfix,longbibliography]{revtex4-2}

\usepackage{color}
\usepackage{mathrsfs}
\usepackage{graphicx}
\usepackage{dcolumn}
\usepackage{bm}
\usepackage{hyperref}
\usepackage{amsmath,amsfonts,amssymb,ulem}
\usepackage{epstopdf}
\usepackage{xcolor}

\newcommand{\re}[1]{{\color{black}#1}}

\usepackage{graphicx}

\begin{document}

\title{Deciphering \textit{in situ} electron dynamics of ultrarelativistic plasma  via \\ polarization pattern of emitted $\gamma$-photons}

\author{Zheng Gong}
\email[]{gong@mpi-hd.mpg.de}
 \affiliation{Max-Planck-Institut f\"{u}r Kernphysik, Saupfercheckweg 1, 69117 Heidelberg, Germany}
\author{Karen Z. Hatsagortsyan}
\email[]{k.hatsagortsyan@mpi-hd.mpg.de}
 \affiliation{Max-Planck-Institut f\"{u}r Kernphysik, Saupfercheckweg 1, 69117 Heidelberg, Germany}
\author{Christoph H. Keitel}
 \affiliation{Max-Planck-Institut f\"{u}r Kernphysik, Saupfercheckweg 1, 69117 Heidelberg, Germany}

\date{\today}

\begin{abstract}

Understanding and interpretation of the dynamics of ultrarelativistic plasma is a challenge, which calls for the development of methods for \textit{in situ} probing the plasma dynamical characteristics. We put forward a new method, harnessing polarization properties of $\gamma$-photons emitted from a non-pre-polarized plasma irradiated by a circularly polarized pulse. We show that the  angular pattern of $\gamma$-photon linear polarization is explicitly correlated with the dynamics of the radiating electrons, which provides information on the laser-plasma interaction regime. Furthermore, with the $\gamma$-photon circular polarization originating from the electron radiative spin-flips, the plasma susceptibility to quantum electrodynamical processes is gauged. Our study demonstrates that the polarization signal of emitted $\gamma$-photons can be a versatile information source, which would be beneficial for the research fields of laser-driven plasma, accelerator science, and laboratory astrophysics.

\end{abstract}

\maketitle
Ultrarelativistic plasma produced at cutting-edge laser facilities with intensities already reaching the level of $10^{23}\,\mathrm{W/cm^2}$~\cite{Mourou_etal_2006,danson2019petawatt,yoon2021realization}, is favorable for investigation of new regimes of electron and ion acceleration~\cite{esarey_LWFA_RMP,macchi_2013_RMP,wang2021super}, and hard photon emissions~\cite{yan2017high,brady2012laser,ridgers2012dense,stark2016enhanced,gonoskov2017ultrabright,magnusson2019laser,zhu2020extremely}, to explore astrophysical phenomena in a laboratory~\cite{extreme_environment,RRD_bulanov2015,gonzalez2016optically}, and to probe nonlinear QED processes in ultrastrong fields \cite{di2012extremely,cole2018experimental,poder2018experimental,qu2021signature,fedeli2021probing}.
The description of plasma in such extreme conditions is a challenge even for numerical particle-in-cell (PIC) simulations, which calls for the development of new methods for \textit{in situ} probing the plasma dynamical characteristics and verification of the employed plasma models.
Unlike underdense plasma being noninvasively diagnosed through optical probes or charged particle radiography~\cite{borghesi2002electric,downer2018diagnostics,bott2021PRL}, the fast evolving ($\sim$fs) ultrarelativistic plasma, associated with sufficiently strong fields~\cite{gonoskov2021charged}, radiative particle trapping~\cite{gonoskov2014anomalous,ji2014radiation,bulanov2017charged}, or $e^-e^+$ pair cascades~\cite{bell2008possibility,nerush2011laser,zhu2016dense}, tends to be energetic and overdense, which hinders the use of conventional diagnostic techniques. New approaches based on XFEL beams~\cite{wang2019structured} or ejected spin-polarized electrons~\cite{gong2021diagnosing} were recently proposed to measure Megatesla-level magnetic fields within overdense plasma. Nevertheless, interrogating and verifying the \textit{in situ} transient electron dynamics remains a challenging covet.

Recently, the successful decoding of field properties nearby a black hole~\cite{BH_image_2021_March} re-stimulated interest in diagnostics based on photon polarization~\cite{lembo2021cosmic}. While polarized light is vulnerable to magneto-optic disturbance~\cite{faraday1846magnetization}, a high-frequency $\gamma$-photon is robust during penetration of the plasma depth~\cite{chen2012introduction}. In contrast to the routinely detected quantities of arrival time and energy, the $\gamma$-photon polarization (GPP), provides new insights into the relativistic jet geometry~\cite{zhang2019detailed} and magnetic field configuration~\cite{gill2020linear}, which allows to identify cosmic neutrino scattering~\cite{batebi2016generation}, dark matter annihilation~\cite{boehm2017circular}, and acceleration mechanisms surrounding crab pulsars~\cite{dean2008polarized}. Prompted by the role of GPP in astrophysical scenarios, the question arises on its possible use to diagnose the \textit{in situ} electron dynamics in ultrarelativistic plasma.

\begin{figure*}[t]
\includegraphics[width=0.9\textwidth]{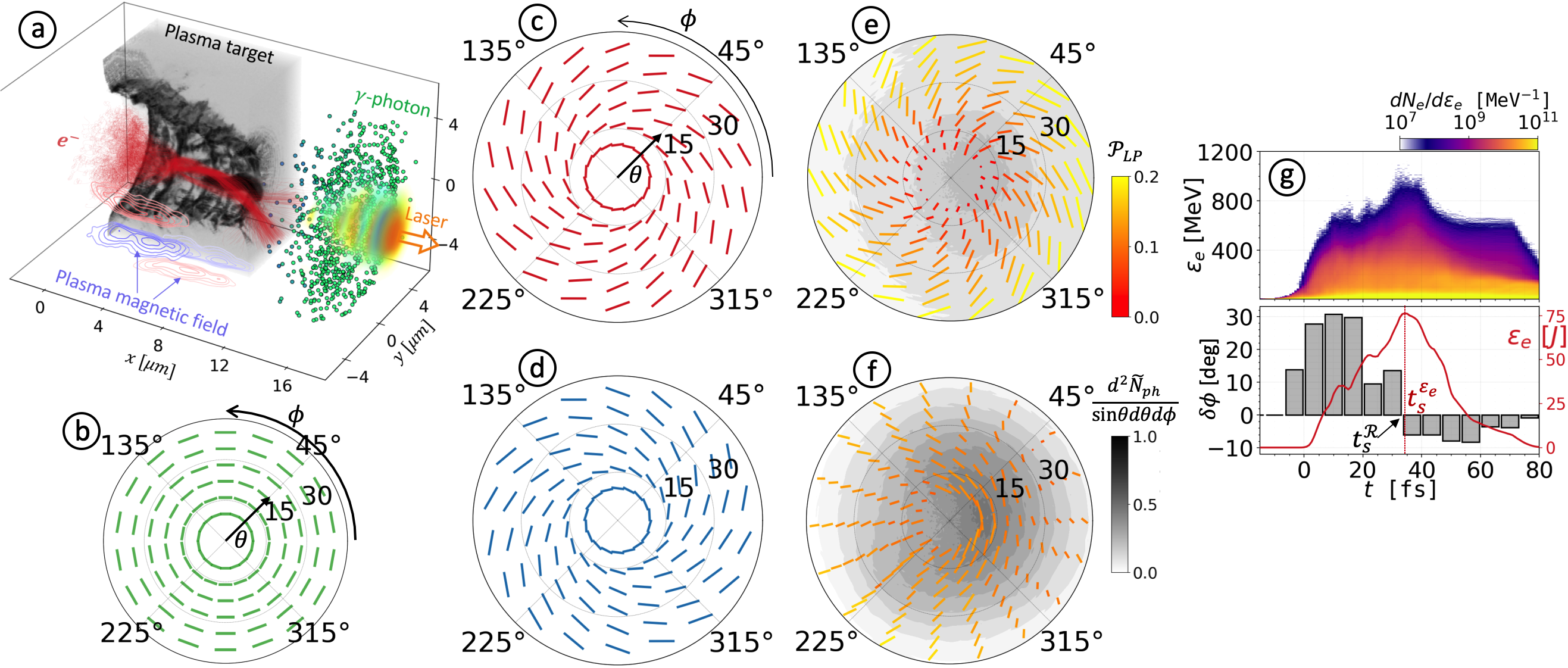}
\caption{(a) The schematic for $\gamma$-photon emissions from a plasma (with $y<0$ clipped) interacting with a laser pulse.
(b) The LP orientation along the local azimuthal direction.
The analytically predicted LP orientation for the electron undergoing acceleration (c) and deceleration (d). The simulated $\gamma$-photon LP for the time at (e) $t=10$ and (f) $40\,$fs, with the LP degree $\mathcal{P}_\mathrm{LP}$ and the normalized number distribution $d^2\widetilde{N}_{ph}/\sin\theta d\theta d\phi$. \re{In (b)-(f), the angular distribution is shown in the space of $(\theta,\phi)$, where $\theta=\arctan[(p_y^2+p_z^2)^{1/2}/p_x]$ and $\phi=\arctan2(p_z,p_y)$ characterise the momentum direction of emitted $\gamma$-photons.}
(g) The time evolution of the electron energy spectrum $dN_e/d\varepsilon_e$, the polarization orientation $\delta\phi$, and the kinetic energy $\mathcal{E}_{e}$ of electrons with $\gamma_e>a_0$.
}
\label{fig:schematic}
\end{figure*}

This Letter aims to find the distinct relationship between the spatial features of the GPP and the time-resolved motion of plasma electrons, and in this way deduce the dynamical properties of plasma and the regime of interaction. Using 3D PIC simulations, we study the polarization-resolved $\gamma$-photon emission in an ultrarelativistic plasma driven by a circularly polarized pulse [see Fig.~\ref{fig:schematic}(a)].
The collective orientation of the $\gamma$-photons' linear polarization (LP) resembles a spiral shape with the rotation tendency determined by the acceleration status of the radiating electrons.
To quantifies the degree of rotation tendency, we introduce the polarization angle $\delta\phi$, as the deviation of the orientation of GPP with respect to the azimuthal direction. It is determined by the electrons' transient acceleration gradient, and its angle dependence can serve as a classifier to distinguish laser-plasma interaction regimes with different electron dynamics, such as with the electrons' single- or multi-cycle resonance oscillations, or the longitudinal braking emission.
Furthermore, the $\gamma$-photons' circular polarization (CP), originating from the accumulated longitudinal polarization of plasma electrons due to the quantum radiative spin-flips, is shown to provide a measure of the susceptibility of the ultrarelativistic plasma to QED processes.

When an electron interacts with an electromagnetic wave $\bm{A}=a_0e^{i\xi}\hat{\bm{e}}_y+\epsilon a_0 e^{i(\xi-\pi/2)}\hat{\bm{e}}_z$ with the normalized amplitude $a_0$, ellipticity $\epsilon=1$, and phase $\xi=\omega_0t-k_0x$ ($\omega_0/k_0=v_{ph}$), the electron motion is described by $p_{y,z}\sim A_{y,z}m_ec$, $y,z\sim -iA_{y,z}/( \Gamma k_0)$, with the dephasing $\Gamma\equiv\gamma_e-(p_x/m_ev_{ph})$.
The laser fields predominantly governs the electron dynamics, while the self-generated azimuthal magnetic field $\overline{\bm{B}}_{\phi}=\kappa_b(-y\hat{\bm{e}}_z+z\hat{\bm{e}}_y)$~\cite{pukhov2002strong} acts as a perturbation. The radially quasi-static electric field is neglected due to the ion motion compensating the charge separation~\cite{gong2020direct}.
The polarized $\gamma$-photon emission and electron radiative spin-flips is determined by the strong-field
quantum parameter $\chi_{e,ph}\equiv (e\hbar/m_e^3c^4) |F_{\mu\nu} p^{\nu}|$ with the field tensor $F_{\mu\nu}$ and the momentum $p^{\nu}$ of the electron or photon, the electron charge $-e $, and mass $m_e$, the Planck constant $\hbar$, and the speed of light $c$. In the moderate QED regime $\chi_e < 1$, the direction of the emitted $\gamma$-photon LP is primarily parallel with the acceleration direction perpendicular to the electron momentum, $\bm{a}_\perp\equiv\bm{a}-(\bm{a}\cdot\hat{\bm{v}}) \hat{\bm{v}}$, where the hat symbol denotes the unit vector. Thus, the polarization orientation can be derived
\begin{eqnarray}\label{eq:a_perp}
\begin{aligned}
\hat{a}_{\perp,y} \approx & -\sin\phi\left(\frac{\Gamma}{\epsilon\gamma_e}+\frac{\kappa_b\cos\theta}{\epsilon\Gamma}\right) -\frac{(-\bm{\beta}\cdot\bm{E})\cos\phi}{\gamma_e},\  \\
\hat{a}_{\perp,z} \approx & \cos\phi\left(\frac{\epsilon\Gamma}{\gamma_e}+\frac{\epsilon\kappa_b\cos\theta}{\Gamma}\right)-\frac{(-\bm{\beta}\cdot\bm{E})\sin\phi}{\gamma_e},
\end{aligned}
\end{eqnarray}where $\bm{\beta}=\bm{v}/c$, $\theta=\arctan[(p_y^2+p_z^2)^{1/2}/p_x]$ and $\phi=\arctan2(p_z,p_y)$.
When the electron energy gain is negligible, i.e. $-\bm{\beta}\cdot\bm{E}=0$, the orientation of the $\gamma$-photon LP would be along the azimuthal direction $\hat{\bm{a}}_a=(-\sin\phi,\cos\phi)^\intercal$, which collectively resembles multiple concentric rings with each polarization segment along the azimuthal direction [Fig.~\ref{fig:schematic}(b)].
The deviation of the $\gamma$-photon LP orientation from the azimuthal direction can be quantified by $\delta\phi\in[-90^\circ,90^\circ]$, which is the relative angle between $\hat{\bm{a}}_\perp$ and $\hat{\bm{a}}_a$~\cite{SM} and is calculated as
\begin{eqnarray}\label{eq:R}
\delta\phi\approx\arcsin\left\{ \frac{-\bm{\beta}\cdot\bm{E}}{\sqrt{\left[\Gamma+(\gamma_e\kappa_b\cos\theta/\Gamma)\right]^2+(\bm{\beta}\cdot\bm{E})^2}}\right\}.
\end{eqnarray}
If the radiating electron is undergoing acceleration with $-\bm{\beta}\cdot\bm{E}>0$ ($-\bm{\beta}\cdot\bm{E}<0$), the GPP orientation $\delta\phi>0$ ($\delta\phi<0$) corresponds to the counter-clockwise (clockwise) spiral tendency in the angular distribution of $\gamma$-photon LP as shown in Fig.~\ref{fig:schematic}(c) [Fig.~\ref{fig:schematic}(d)].

To examine the GPP features, we performed 3D PIC simulations, where a non-pre-polarized slab is illuminated by a circularly polarized pulse ($\epsilon=1$). The laser intensity  $I_0\approx 1.7 \times 10^{23}\mathrm{W/cm}^2$ is equivalent to $a_0 \approx 350$ for the wavelength $\lambda_0=1\,\mu m$. The pulse has a duration $\tau_0=25\,$~fs and focal spot size $2.6\,\mu$m (FWHM intensity measure). The plasma slab has a thickness $l_0=10\,\mu$m and consists of electrons and carbon ions with the density $n_e=30n_c$ and $n_i=5n_c$, respectively, $n_{c}= m_e \omega_0^2 / 4 \pi e^2$ is the plasma critical density. The influences of radiative spin-flips, spin-dependent photon emission, and photon polarization effects have been incorporated in the EPOCH code~\cite{arber2015contemporary,SM}.

\begin{figure}
\includegraphics[width=1\columnwidth]{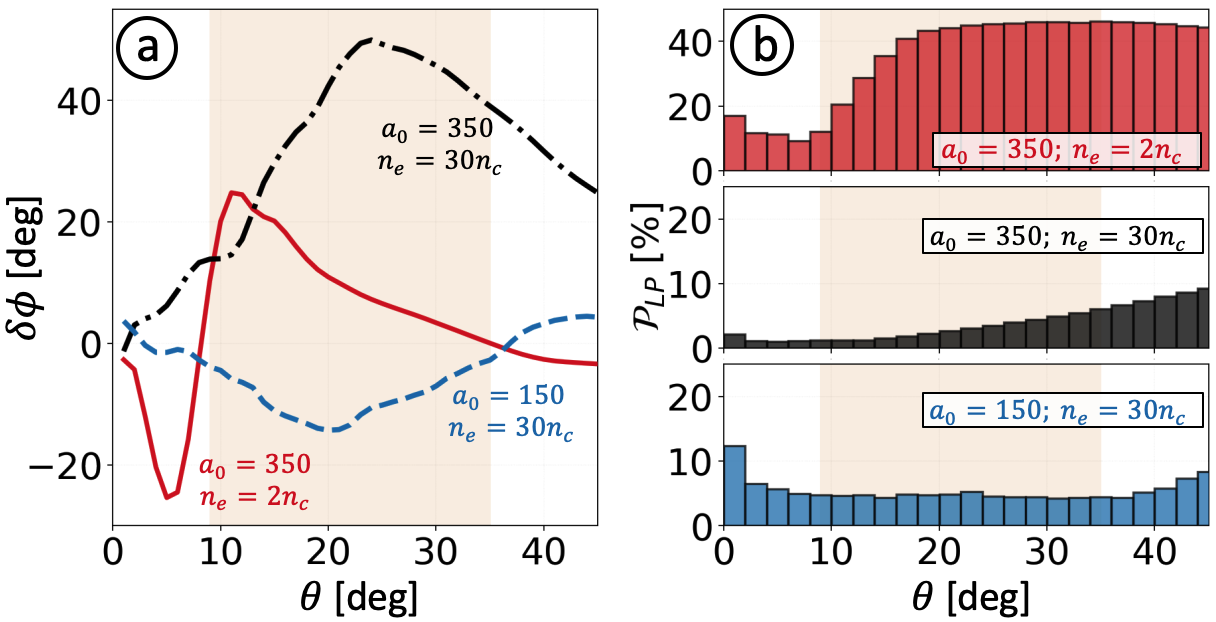}
\caption{The angular distribution of (a) polarization angle $\delta\phi(\theta)$ and (b) LP degree $\mathcal{P}_{LP}(\theta)$ for three distinct regimes of SRO (red), MRO (black), and LBE (blue). The shadow area is marked for comparison with Fig.~\ref{fig:one_cycle}(a) and Fig.~\ref{fig:multi_cycle}(a).}
\label{fig:regimes}
\end{figure}

Inside the interacting plasma, the electrons tend to form a helical density structure~\cite{liu2013generating}, undergo betatron oscillation~\cite{pukhov1999_DLA}, and radiate multi-MeV photons~\cite{liu2015quasimonoenergetic}. The orientation of the emitted $\gamma$-photon LP exhibits a counter-clockwise spiral tendency at $t=10$\,fs [Fig.~\ref{fig:schematic}(e)] reproducing well the analytical prediction for the accelerating electron. Here, the averaged polarization angle is $\delta\phi\approx 30.6^\circ$ and the LP degree $\mathcal{P}_\mathrm{LP}\approx 15.3\%$. The clockwise spiral tendency in Fig.~\ref{fig:schematic}(f) implies the deceleration of plasma electrons occurring later at $t=40\,$fs.  The time-resolved $\delta\phi$ explicitly reflects the electrons being predominantly accelerated (decelerated) at $t\lesssim t_s^{\varepsilon_e}$ ($t\gtrsim t_s^{\varepsilon_e}$) [see Fig.~\ref{fig:schematic}(g)], where $t_s^{\varepsilon_e}\sim 35\,$fs is the electron energy saturation time. Consequently, the moment of $\delta\phi$ changing sign, defined as the reversal time $t_s^\mathcal{R}$, should be equal to $t_s^{\varepsilon_e}$, which is confirmed by simulations for different parameters~\cite{SM}. Although the measurement of $t_s^\mathcal{R}$ can provide time-resolved plasma information, the timing accuracy of $\sim10\,$fs is not available yet for the current $\gamma$-photon polarimetry~\cite{tatischeff2017astrogam}. Therefore, we turn to the investigation of the angle-dependence of $\delta\phi(\theta)$ of all emitted $\gamma$-photons during the interaction [Fig.~\ref{fig:regimes}].

Three distinct typical angle distributions of $\delta\phi(\theta)$ are possible when varying $a_0$ and $n_e$ [Fig.~\ref{fig:regimes}(a)]. Each typical distribution corresponds to a specific  category of the electron dynamics: the single-cycle resonance oscillation (SRO) regime [$a_0$$=$$350$; $n_e$$=$$2n_c$], the multi-cycle resonance oscillation (MRO) regime  [$a_0$$=$$350$; $n_e$$=$$30n_c$], and the longitudinal braking emission (LBE) regime at weak fields and high density [$a_0$$=$$150$; $n_e$$=$$30n_c$].

\begin{figure}
\includegraphics[width=1\columnwidth]{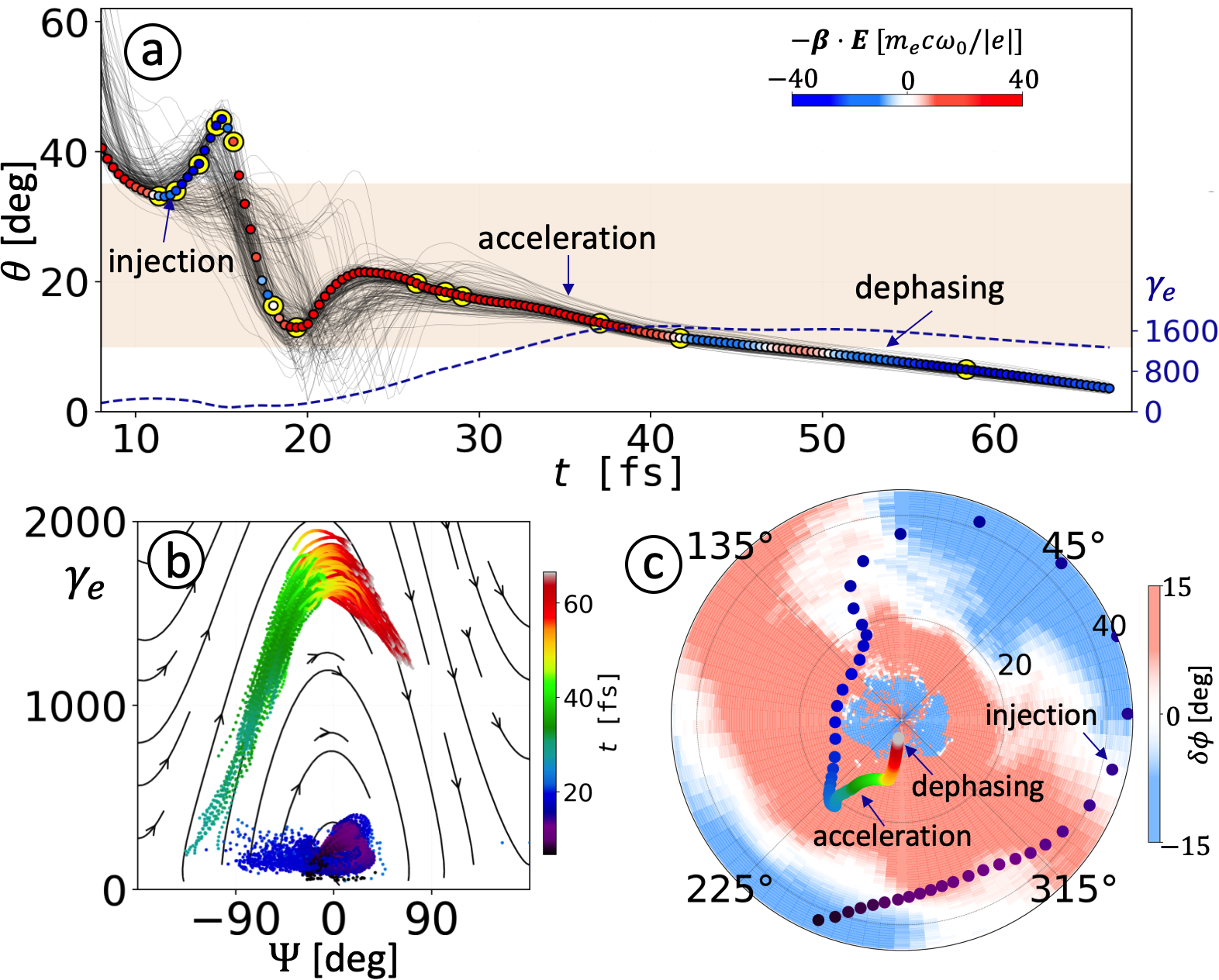}
\caption{The electron dynamics of the SRO regime.
(a) The time evolution of the electron's angle $\theta$ with blue-red color denoting $-\bm{\beta}\cdot\bm{E}$ and $\gamma$-photon emissions marked by yellow circles. The thin black lines represent different electrons. The blue dashed line refers to $\gamma_e$. (b) The evolution of the representative electrons in $(\Psi,\gamma_e)$ space. \re{(c) The evolution of the electron in $(\theta,\phi)$ space, where the background color manifests the detected $\delta\phi(\theta,\phi)$. The rainbow color in (b)(c) refers to the time.}}
\label{fig:one_cycle}
\end{figure}

\textit{Single-cycle resonance oscillation} (at strong fields and low densities)  -- the typical electron experiences an angle-dependent acceleration and deceleration procedure after being injected into one laser period [Fig.~\ref{fig:one_cycle}(a)]. This injection is triggered by the suppressed dephasing $\Gamma\sim\int (1-\beta_x)E_xdt$ due to a negative $E_x$ exerted on the electron. Here, the oscillating laser field $E_x\sim(\lambda_0/2\pi)\partial E_r/\partial r\approx 10^{14}\mathrm{V/m}$ rather than the quasistatic self-generated field is the dominant term of the whole longitudinal electric field. When the electron moves in a range with negative $E_x$ at $6<t<12\,\mathrm{fs}$, its dephasing value decreases from $\Gamma/\gamma_e\approx1.4$ to $0.2$. If the relative angle between the laser vector potential and electron's transverse momentum is defined as $\Psi:=\xi-\phi$, the electron dynamics can be described by $d\gamma_e/dt=-a_0\sin\theta\sin\Psi -\beta_xE_x -\bm{\beta}\cdot\bm{f}_{RR}$ and $d\Psi/dt=\omega_0(1-v_x/v_{ph})-\omega_\beta$ in $(\Psi,\gamma_e)$ space, where $\bm{f}_{RR}$ is the radiation reaction force and $\omega_\beta\approx\sqrt{|e|v_x\kappa_b/\gamma_em_e}$. The time evolution of the Lorentz factor $\gamma_e$ [Fig.~\ref{fig:one_cycle}(a)] manifests that the interacting scenario takes place merely in one laser period at $20<t<60\,$fs, which is further confirmed by the electron evolution in the space of $(\Psi,\gamma_e)$ [Fig.~\ref{fig:one_cycle}(b)].
As illustrated in Fig.~\ref{fig:one_cycle}(c), features of $\delta\phi|_{\theta>35^\circ}$$<$$0$, $\delta\phi|_{10<\theta<35^\circ}$$>$$0$, and $\delta\phi|_{\theta<10^\circ}$$<$$0$ correspond to the electron's injection termination, phase matching acceleration, and dephasing deceleration, respectively. The panel (b) in Fig.~\ref{fig:one_cycle} indicates that the electron dynamics in the plasma channel is quasi-synchronous during a single resonance oscillation,
leading to strong coherence of the emitted $\gamma$-photons. Thus, the LP degree of the SRO regime $\mathcal{P}_{LP}\approx40.7\%$ is approximately one order of magnitude higher than those of the MRO ($3.2\%$) and LBE ($5.3\%$) regimes [see Fig.~\ref{fig:regimes}(b)].

\begin{figure}[t]
\includegraphics[width=1\columnwidth]{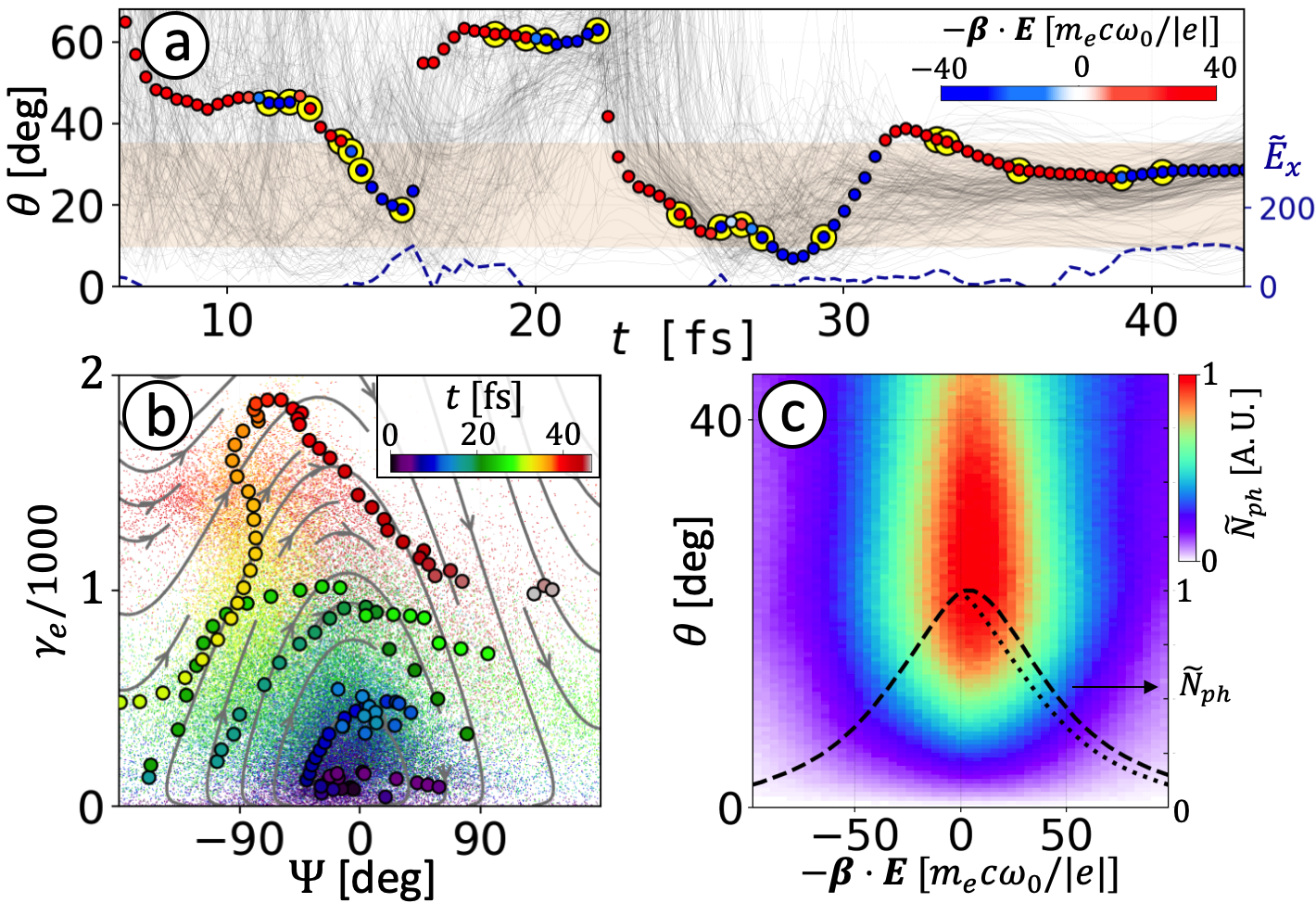}
\caption{The electron dynamics of the MRO regime. (a) The same as Fig.~\ref{fig:one_cycle}(a) but for the MRO regime, where the dashed blue line denotes the normalized field $\widetilde{E}_x=|e|E_x/m_ec\omega_0$. (b) The single electron trajectory (circles) and evolution tendency of a bunch of electrons (small dots) in $(\Psi,\gamma_e)$ space, where the rainbow color refers to time. (c) The dependence of the $\gamma$-photon number distribution $\widetilde{N}_{ph}$ on $-\bm{v}\cdot\bm{E}$ and $\theta$, where the dashed line shows $\widetilde{N}_{ph}(-\bm{v}\cdot\bm{E})$ and the dotted line displays the symmetry of $\widetilde{N}_{ph}(-\bm{v}\cdot\bm{E}<0)$ at $-\bm{v}\cdot\bm{E}>0$ for comparison.}
\label{fig:multi_cycle}
\end{figure}

\textit{Multi-cycle resonance oscillation} (at strong fields and high densities) -- as the gradient of plasma magnetic field $\kappa_b\sim (m_e\omega_0^2/|e|c)(n_e/n_c)$~\cite{stark2016enhanced} is enhanced due to the raised density, the increased oscillation frequency $\omega_\beta\propto n_e^{1/2}$ readily mismatches with the relatively laser frequency $d\xi/dt$.  Consequently, the electron, associated with the abrupt change of $\theta$ by the imposed field $E_x$, repeatedly experiences dephasing and slides into the next accelerating phase [see Fig.~\ref{fig:multi_cycle}(a)], which is further verified by its multiple rotation in $(\Psi,\gamma_e)$ space [Fig.~\ref{fig:multi_cycle}(b)]. Although the MRO scenario can be qualitatively decomposed into multiple single-cycle resonance oscillations, 
the multiple injections with varying locations
and fast dephasing deteriorate the synchronous motion [Fig.~\ref{fig:multi_cycle}(a)], which gives rise to a diffusive distribution of electron evolution in $(\Psi,\gamma_e)$ phase space [Fig.~\ref{fig:multi_cycle}(b)] and a low LP degree $\mathcal{P}_{LP}\approx3.2\%$ of emitted $\gamma$-photons. As shown in Fig.~\ref{fig:multi_cycle}(c), the dependence of photon number distribution $\widetilde{N}_{ph}$ on angle $\theta$ and acceleration status $-\bm{v}\cdot\bm{E}$ demonstrates that overall the acceleration is dominant in the electron's energy exchange with the laser field, which agrees with the distribution of $\delta\phi(\theta<40^\circ)>0$.


\begin{figure}[t]
\includegraphics[width=0.9\columnwidth]{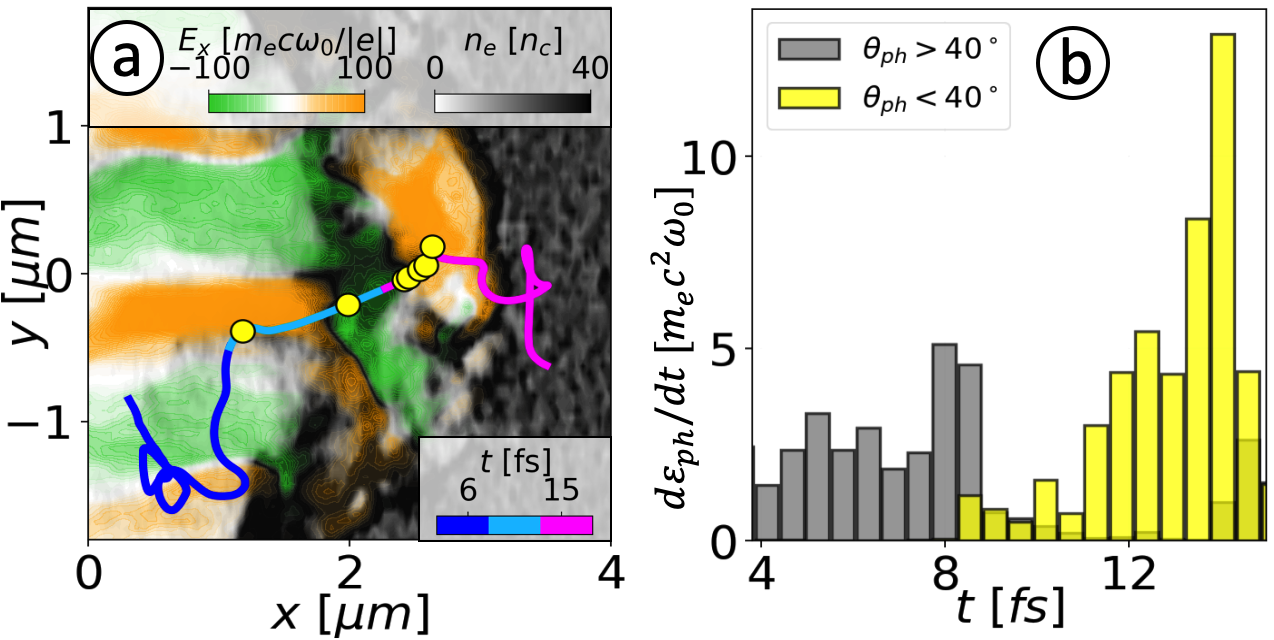}
\caption{The electron dynamics of the LBE regime. \re{(a) The representative electron trajectory in $(x,y)$ plane}, where the yellow circles denote the emitted $\gamma$-photons with energy $\varepsilon_{ph}>10m_ec^2$. (b) The time evolution of the photon emission power $d\varepsilon_{ph}/dt$.}
\label{fig:lbe}
\end{figure}

\textit{Longitudinal braking emission} (at relatively weak fields and high densities) -- If a strong charge separation sustained at the interface between the laser wave front and the plasma channel edge, the electron channeling into the place is hampered. The electron is primarily braked and emits $\gamma$-photons within a collimated polar angle $\theta$, while being exposed to the positive longitudinal electric field $E_x$ \re{[Fig.~\ref{fig:lbe}(a)(b)]}. In the braking emission stage, the term of $-v_{\parallel}E_{\parallel}$ ($-v_{\perp}E_{\perp}$) contributes $86\%$ ($14\%$) of the whole $-\bm{v}\cdot\bm{E}$, and the averaged orientation of GPP is $\delta\phi\approx -8.9^\circ$. Therefore, the $\delta\phi(\theta)<0$ holds over a large range of $\theta<40^\circ$.
The braking dynamics develops as follows. The positive $E_x$ moves forward with a velocity $v_m\sim0.32 c$, where the new replenished electrons with a near-light velocity $v_x \gtrsim 0.75c$ quickly catch up with the moving $E_x$ and then get braked there. The polarization pattern in the $(\theta,\phi)$ space for the LBE regime is distinct [cf. Fig.~S6 in \cite{SM} with Fig.~\ref{fig:one_cycle}(c)] determined by the main cycle of the laser pulse and is CEP dependent.

\begin{figure}[t]
\includegraphics[width=1\columnwidth]{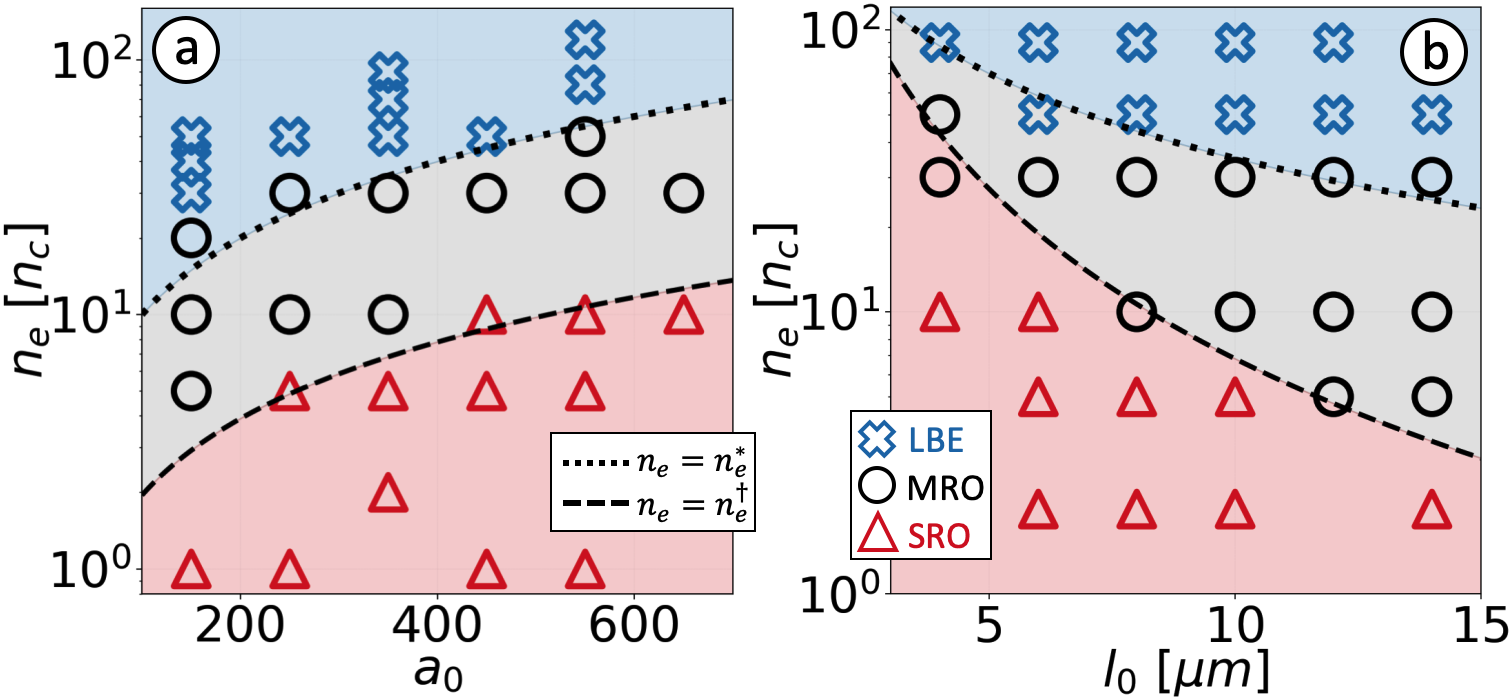}
\caption{The range of the different regimes presented in the parameter space of $a_0$ and $n_e$: (a) for $l_0=10\,\mu $m and (b) for $a_0=350$. The markers denote the regimes of SRO (triangles), MRO (circles), and LBE (crosses) identified by the simulation results of $\delta\phi(\theta)$. The dashed and dotted lines refer to $n_e^\dagger$ and $n_e^*$, respectively.}
\label{fig:scan}
\end{figure}

In Fig.~\ref{fig:scan} we present the range of the different regimes  in the parameter space of $a_0$ and $n_e$, which can be identified by the $\gamma$-photon polarization. The criterion of separation of the SRO and MRO regimes is estimated from the following condition that the electron undergoes one period of resonant oscillation during the interaction: $\omega_\beta(l_0/c)\sim 2\pi$, which is reformulated as $n_e^\dagger\sim (\pi m_ec^2/e^2)a_0l_0^{-2}$ [Fig.~\ref{fig:scan}].

The threshold of LBE is estimated as follows. The LBE stems from the intense quasistatic longitudinal electric field, which is sustained when the laser pulse is readily depleted within the plasma slab: $(v_{ph}-v_g)l_0/v_{ph}\gtrsim \tau_0c$, where $v_{ph}\sim c/\sqrt{1-n_e/(a_0n_c)}$ and $v_{g}=c^2/v_{ph}$. The latter conditions give the threshold density for LBE: $n_e\gtrsim n_e^*\sim (m_ec\omega_0/2e^2)a_0l_0^{-1}$.


Besides the LP, $\gamma$-photons are partially circularly polarized (CP), see \cite{SM}, which originates from the spin polarization of  plasma electrons, caused by electron radiative spin flips. The latter are governed by the QED quantum strong-field parameter $\chi_e$. Thus, the CP of emitted $\gamma$-photons can be a characteristic of the QED properties of the laser-driven plasma. We provide analytical estimation of the CP degree in~\cite{SM} and show its agreement with PIC simulation results.


Concluding, we demonstrate that the polarization properties of $\gamma$-photons emitted from an ultrarelativistic plasma driven by a circularly polarized pulse provide exclusive information on the \textit{in situ} electron dynamics, which allows to identify three specific regimes of laser-plasma interaction, as well as the plasma QED status. Our diagnostic scheme via the emitted $\gamma$-photon polarization is mainly applied in the laser-driven ultrarelativistic overdense plasma, which cannot be measured by the conventional optical probes. In addition,
since the ultrarelativistic plasma is generally associated
with sufficiently strong fields, the requirement of the ion
energy on the proton radiography is unfeasible either.
Therefore, the advantage of the diagnostic based on $\gamma$-photon polarization is its applicability in the energetic
and overdense plasma. \re{Note that the requirement of the angular resolution ($\sim1^\circ$) of GPP is satisfied by the gammaray polarimetry parameters envisaged in Ref.~\cite{rokujo2018first}, and thus the method presented here is likely to be experimentally feasible in the near future.} 
This diagnostic tool may appear beneficial for better understanding of phenomena in broad high-intensity interaction scenarios including ion acceleration~\cite{qiao2009stable,chen2010radiation,tamburini2010radiation,mcilvenny2021selective}, direct laser acceleration~\cite{pukhov1999_DLA}, high-harmonic generation~\cite{wang2019intense,yi2021high}, brilliant photon emission~\cite{yi2016bright,wang2018collimated}, ultradense nanopinches~\cite{kaymak2016nanoscale}, and $e^-e^+$ pair plasma cascades~\cite{grismayer2016laser}.


The PIC code EPOCH is funded by the UK EPSRC grants EP/G054950/1, EP/G056803/1, EP/G055165/1 and EP/ M022463/1. Z. G. would like to thank Pei-Lun He for fruitful discussions.



\begin{thebibliography}{74}%
\makeatletter
\providecommand \@ifxundefined [1]{%
 \@ifx{#1\undefined}
}%
\providecommand \@ifnum [1]{%
 \ifnum #1\expandafter \@firstoftwo
 \else \expandafter \@secondoftwo
 \fi
}%
\providecommand \@ifx [1]{%
 \ifx #1\expandafter \@firstoftwo
 \else \expandafter \@secondoftwo
 \fi
}%
\providecommand \natexlab [1]{#1}%
\providecommand \enquote  [1]{``#1''}%
\providecommand \bibnamefont  [1]{#1}%
\providecommand \bibfnamefont [1]{#1}%
\providecommand \citenamefont [1]{#1}%
\providecommand \href@noop [0]{\@secondoftwo}%
\providecommand \href [0]{\begingroup \@sanitize@url \@href}%
\providecommand \@href[1]{\@@startlink{#1}\@@href}%
\providecommand \@@href[1]{\endgroup#1\@@endlink}%
\providecommand \@sanitize@url [0]{\catcode `\\12\catcode `\$12\catcode
  `\&12\catcode `\#12\catcode `\^12\catcode `\_12\catcode `\%12\relax}%
\providecommand \@@startlink[1]{}%
\providecommand \@@endlink[0]{}%
\providecommand \url  [0]{\begingroup\@sanitize@url \@url }%
\providecommand \@url [1]{\endgroup\@href {#1}{\urlprefix }}%
\providecommand \urlprefix  [0]{URL }%
\providecommand \Eprint [0]{\href }%
\providecommand \doibase [0]{https://doi.org/}%
\providecommand \selectlanguage [0]{\@gobble}%
\providecommand \bibinfo  [0]{\@secondoftwo}%
\providecommand \bibfield  [0]{\@secondoftwo}%
\providecommand \translation [1]{[#1]}%
\providecommand \BibitemOpen [0]{}%
\providecommand \bibitemStop [0]{}%
\providecommand \bibitemNoStop [0]{.\EOS\space}%
\providecommand \EOS [0]{\spacefactor3000\relax}%
\providecommand \BibitemShut  [1]{\csname bibitem#1\endcsname}%
\let\auto@bib@innerbib\@empty
\bibitem [{\citenamefont {Mourou}\ \emph {et~al.}(2006)\citenamefont {Mourou},
  \citenamefont {Tajima},\ and\ \citenamefont {Bulanov}}]{Mourou_etal_2006}%
  \BibitemOpen
  \bibfield  {author} {\bibinfo {author} {\bibfnamefont {G.~A.}\ \bibnamefont
  {Mourou}}, \bibinfo {author} {\bibfnamefont {T.}~\bibnamefont {Tajima}},\
  and\ \bibinfo {author} {\bibfnamefont {S.~V.}\ \bibnamefont {Bulanov}},\
  }\bibfield  {title} {\bibinfo {title} {Optics in the relativistic regime},\
  }\href@noop {} {\bibfield  {journal} {\bibinfo  {journal} {Rev. Mod. Phys.}\
  }\textbf {\bibinfo {volume} {78}},\ \bibinfo {pages} {309} (\bibinfo {year}
  {2006})}\BibitemShut {NoStop}%
\bibitem [{\citenamefont {Danson}\ \emph {et~al.}(2019)\citenamefont {Danson}
  \emph {et~al.}}]{danson2019petawatt}%
  \BibitemOpen
  \bibfield  {author} {\bibinfo {author} {\bibfnamefont {C.~N.}\ \bibnamefont
  {Danson}} \emph {et~al.},\ }\bibfield  {title} {\bibinfo {title} {Petawatt
  and exawatt class lasers worldwide},\ }\href@noop {} {\bibfield  {journal}
  {\bibinfo  {journal} {High Power Laser Science and Engineering}\ }\textbf
  {\bibinfo {volume} {7}} (\bibinfo {year} {2019})}\BibitemShut {NoStop}%
\bibitem [{\citenamefont {Yoon}\ \emph {et~al.}(2021)\citenamefont {Yoon} \emph
  {et~al.}}]{yoon2021realization}%
  \BibitemOpen
  \bibfield  {author} {\bibinfo {author} {\bibfnamefont {J.~W.}\ \bibnamefont
  {Yoon}} \emph {et~al.},\ }\bibfield  {title} {\bibinfo {title} {Realization
  of laser intensity over $10^{23}$ w/cm$^2$},\ }\href@noop {} {\bibfield
  {journal} {\bibinfo  {journal} {Optica}\ }\textbf {\bibinfo {volume} {8}},\
  \bibinfo {pages} {630} (\bibinfo {year} {2021})}\BibitemShut {NoStop}%
\bibitem [{\citenamefont {Esarey}\ \emph {et~al.}(2009)\citenamefont {Esarey},
  \citenamefont {Schroeder},\ and\ \citenamefont {Leemans}}]{esarey_LWFA_RMP}%
  \BibitemOpen
  \bibfield  {author} {\bibinfo {author} {\bibfnamefont {E.}~\bibnamefont
  {Esarey}}, \bibinfo {author} {\bibfnamefont {C.}~\bibnamefont {Schroeder}},\
  and\ \bibinfo {author} {\bibfnamefont {W.}~\bibnamefont {Leemans}},\
  }\bibfield  {title} {\bibinfo {title} {Physics of laser-driven plasma-based
  electron accelerators},\ }\href@noop {} {\bibfield  {journal} {\bibinfo
  {journal} {Reviews of Modern Physics}\ }\textbf {\bibinfo {volume} {81}},\
  \bibinfo {pages} {1229} (\bibinfo {year} {2009})}\BibitemShut {NoStop}%
\bibitem [{\citenamefont {Macchi}\ \emph {et~al.}(2013)\citenamefont {Macchi},
  \citenamefont {Borghesi},\ and\ \citenamefont {Passoni}}]{macchi_2013_RMP}%
  \BibitemOpen
  \bibfield  {author} {\bibinfo {author} {\bibfnamefont {A.}~\bibnamefont
  {Macchi}}, \bibinfo {author} {\bibfnamefont {M.}~\bibnamefont {Borghesi}},\
  and\ \bibinfo {author} {\bibfnamefont {M.}~\bibnamefont {Passoni}},\
  }\bibfield  {title} {\bibinfo {title} {Ion acceleration by superintense
  laser-plasma interaction},\ }\href@noop {} {\bibfield  {journal} {\bibinfo
  {journal} {Reviews of Modern Physics}\ }\textbf {\bibinfo {volume} {85}},\
  \bibinfo {pages} {751} (\bibinfo {year} {2013})}\BibitemShut {NoStop}%
\bibitem [{\citenamefont {Wang}\ \emph {et~al.}(2021)\citenamefont {Wang} \emph
  {et~al.}}]{wang2021super}%
  \BibitemOpen
  \bibfield  {author} {\bibinfo {author} {\bibfnamefont {P.}~\bibnamefont
  {Wang}} \emph {et~al.},\ }\bibfield  {title} {\bibinfo {title} {Super-heavy
  ions acceleration driven by ultrashort laser pulses at ultrahigh intensity},\
  }\href@noop {} {\bibfield  {journal} {\bibinfo  {journal} {Physical Review
  X}\ }\textbf {\bibinfo {volume} {11}},\ \bibinfo {pages} {021049} (\bibinfo
  {year} {2021})}\BibitemShut {NoStop}%
\bibitem [{\citenamefont {Yan}\ \emph {et~al.}(2017)\citenamefont {Yan},
  \citenamefont {Fruhling}, \citenamefont {Golovin}, \citenamefont {Haden},
  \citenamefont {Luo}, \citenamefont {Zhang}, \citenamefont {Zhao},
  \citenamefont {Zhang}, \citenamefont {Liu}, \citenamefont {Chen} \emph
  {et~al.}}]{yan2017high}%
  \BibitemOpen
  \bibfield  {author} {\bibinfo {author} {\bibfnamefont {W.}~\bibnamefont
  {Yan}}, \bibinfo {author} {\bibfnamefont {C.}~\bibnamefont {Fruhling}},
  \bibinfo {author} {\bibfnamefont {G.}~\bibnamefont {Golovin}}, \bibinfo
  {author} {\bibfnamefont {D.}~\bibnamefont {Haden}}, \bibinfo {author}
  {\bibfnamefont {J.}~\bibnamefont {Luo}}, \bibinfo {author} {\bibfnamefont
  {P.}~\bibnamefont {Zhang}}, \bibinfo {author} {\bibfnamefont
  {B.}~\bibnamefont {Zhao}}, \bibinfo {author} {\bibfnamefont {J.}~\bibnamefont
  {Zhang}}, \bibinfo {author} {\bibfnamefont {C.}~\bibnamefont {Liu}}, \bibinfo
  {author} {\bibfnamefont {M.}~\bibnamefont {Chen}}, \emph {et~al.},\
  }\bibfield  {title} {\bibinfo {title} {High-order multiphoton thomson
  scattering},\ }\href@noop {} {\bibfield  {journal} {\bibinfo  {journal}
  {Nature Photonics}\ }\textbf {\bibinfo {volume} {11}},\ \bibinfo {pages}
  {514} (\bibinfo {year} {2017})}\BibitemShut {NoStop}%
\bibitem [{\citenamefont {Brady}\ \emph {et~al.}(2012)\citenamefont {Brady},
  \citenamefont {Ridgers}, \citenamefont {Arber}, \citenamefont {Bell},\ and\
  \citenamefont {Kirk}}]{brady2012laser}%
  \BibitemOpen
  \bibfield  {author} {\bibinfo {author} {\bibfnamefont {C.~S.}\ \bibnamefont
  {Brady}}, \bibinfo {author} {\bibfnamefont {C.}~\bibnamefont {Ridgers}},
  \bibinfo {author} {\bibfnamefont {T.}~\bibnamefont {Arber}}, \bibinfo
  {author} {\bibfnamefont {A.}~\bibnamefont {Bell}},\ and\ \bibinfo {author}
  {\bibfnamefont {J.}~\bibnamefont {Kirk}},\ }\bibfield  {title} {\bibinfo
  {title} {Laser absorption in relativistically underdense plasmas by
  synchrotron radiation},\ }\href@noop {} {\bibfield  {journal} {\bibinfo
  {journal} {Physical review letters}\ }\textbf {\bibinfo {volume} {109}},\
  \bibinfo {pages} {245006} (\bibinfo {year} {2012})}\BibitemShut {NoStop}%
\bibitem [{\citenamefont {Ridgers}\ \emph {et~al.}(2012)\citenamefont
  {Ridgers}, \citenamefont {Brady}, \citenamefont {Duclous}, \citenamefont
  {Kirk}, \citenamefont {Bennett}, \citenamefont {Arber}, \citenamefont
  {Robinson},\ and\ \citenamefont {Bell}}]{ridgers2012dense}%
  \BibitemOpen
  \bibfield  {author} {\bibinfo {author} {\bibfnamefont {C.}~\bibnamefont
  {Ridgers}}, \bibinfo {author} {\bibfnamefont {C.~S.}\ \bibnamefont {Brady}},
  \bibinfo {author} {\bibfnamefont {R.}~\bibnamefont {Duclous}}, \bibinfo
  {author} {\bibfnamefont {J.}~\bibnamefont {Kirk}}, \bibinfo {author}
  {\bibfnamefont {K.}~\bibnamefont {Bennett}}, \bibinfo {author} {\bibfnamefont
  {T.}~\bibnamefont {Arber}}, \bibinfo {author} {\bibfnamefont
  {A.}~\bibnamefont {Robinson}},\ and\ \bibinfo {author} {\bibfnamefont
  {A.}~\bibnamefont {Bell}},\ }\bibfield  {title} {\bibinfo {title} {Dense
  electron-positron plasmas and ultraintense $\gamma$ rays from
  laser-irradiated solids},\ }\href@noop {} {\bibfield  {journal} {\bibinfo
  {journal} {Physical review letters}\ }\textbf {\bibinfo {volume} {108}},\
  \bibinfo {pages} {165006} (\bibinfo {year} {2012})}\BibitemShut {NoStop}%
\bibitem [{\citenamefont {Stark}\ \emph {et~al.}(2016)\citenamefont {Stark},
  \citenamefont {Toncian},\ and\ \citenamefont {Arefiev}}]{stark2016enhanced}%
  \BibitemOpen
  \bibfield  {author} {\bibinfo {author} {\bibfnamefont {D.}~\bibnamefont
  {Stark}}, \bibinfo {author} {\bibfnamefont {T.}~\bibnamefont {Toncian}},\
  and\ \bibinfo {author} {\bibfnamefont {A.}~\bibnamefont {Arefiev}},\
  }\bibfield  {title} {\bibinfo {title} {Enhanced multi-mev photon emission by
  a laser-driven electron beam in a self-generated magnetic field},\
  }\href@noop {} {\bibfield  {journal} {\bibinfo  {journal} {Physical review
  letters}\ }\textbf {\bibinfo {volume} {116}},\ \bibinfo {pages} {185003}
  (\bibinfo {year} {2016})}\BibitemShut {NoStop}%
\bibitem [{\citenamefont {Gonoskov}\ \emph {et~al.}(2017)\citenamefont
  {Gonoskov}, \citenamefont {Bashinov}, \citenamefont {Bastrakov},
  \citenamefont {Efimenko}, \citenamefont {Ilderton}, \citenamefont {Kim},
  \citenamefont {Marklund}, \citenamefont {Meyerov}, \citenamefont {Muraviev},\
  and\ \citenamefont {Sergeev}}]{gonoskov2017ultrabright}%
  \BibitemOpen
  \bibfield  {author} {\bibinfo {author} {\bibfnamefont {A.}~\bibnamefont
  {Gonoskov}}, \bibinfo {author} {\bibfnamefont {A.}~\bibnamefont {Bashinov}},
  \bibinfo {author} {\bibfnamefont {S.}~\bibnamefont {Bastrakov}}, \bibinfo
  {author} {\bibfnamefont {E.}~\bibnamefont {Efimenko}}, \bibinfo {author}
  {\bibfnamefont {A.}~\bibnamefont {Ilderton}}, \bibinfo {author}
  {\bibfnamefont {A.}~\bibnamefont {Kim}}, \bibinfo {author} {\bibfnamefont
  {M.}~\bibnamefont {Marklund}}, \bibinfo {author} {\bibfnamefont
  {I.}~\bibnamefont {Meyerov}}, \bibinfo {author} {\bibfnamefont
  {A.}~\bibnamefont {Muraviev}},\ and\ \bibinfo {author} {\bibfnamefont
  {A.}~\bibnamefont {Sergeev}},\ }\bibfield  {title} {\bibinfo {title}
  {Ultrabright gev photon source via controlled electromagnetic cascades in
  laser-dipole waves},\ }\href@noop {} {\bibfield  {journal} {\bibinfo
  {journal} {Physical Review X}\ }\textbf {\bibinfo {volume} {7}},\ \bibinfo
  {pages} {041003} (\bibinfo {year} {2017})}\BibitemShut {NoStop}%
\bibitem [{\citenamefont {Magnusson}\ \emph {et~al.}(2019)\citenamefont
  {Magnusson}, \citenamefont {Gonoskov}, \citenamefont {Marklund},
  \citenamefont {Esirkepov}, \citenamefont {Koga}, \citenamefont {Kondo},
  \citenamefont {Kando}, \citenamefont {Bulanov}, \citenamefont {Korn},\ and\
  \citenamefont {Bulanov}}]{magnusson2019laser}%
  \BibitemOpen
  \bibfield  {author} {\bibinfo {author} {\bibfnamefont {J.}~\bibnamefont
  {Magnusson}}, \bibinfo {author} {\bibfnamefont {A.}~\bibnamefont {Gonoskov}},
  \bibinfo {author} {\bibfnamefont {M.}~\bibnamefont {Marklund}}, \bibinfo
  {author} {\bibfnamefont {T.~Z.}\ \bibnamefont {Esirkepov}}, \bibinfo {author}
  {\bibfnamefont {J.}~\bibnamefont {Koga}}, \bibinfo {author} {\bibfnamefont
  {K.}~\bibnamefont {Kondo}}, \bibinfo {author} {\bibfnamefont
  {M.}~\bibnamefont {Kando}}, \bibinfo {author} {\bibfnamefont
  {S.}~\bibnamefont {Bulanov}}, \bibinfo {author} {\bibfnamefont
  {G.}~\bibnamefont {Korn}},\ and\ \bibinfo {author} {\bibfnamefont
  {S.}~\bibnamefont {Bulanov}},\ }\bibfield  {title} {\bibinfo {title}
  {Laser-particle collider for multi-gev photon production},\ }\href@noop {}
  {\bibfield  {journal} {\bibinfo  {journal} {Physical review letters}\
  }\textbf {\bibinfo {volume} {122}},\ \bibinfo {pages} {254801} (\bibinfo
  {year} {2019})}\BibitemShut {NoStop}%
\bibitem [{\citenamefont {Zhu}\ \emph {et~al.}(2020)\citenamefont {Zhu},
  \citenamefont {Chen}, \citenamefont {Weng}, \citenamefont {Yu}, \citenamefont
  {Wang}, \citenamefont {He}, \citenamefont {Sheng}, \citenamefont {McKenna},
  \citenamefont {Jaroszynski},\ and\ \citenamefont {Zhang}}]{zhu2020extremely}%
  \BibitemOpen
  \bibfield  {author} {\bibinfo {author} {\bibfnamefont {X.-L.}\ \bibnamefont
  {Zhu}}, \bibinfo {author} {\bibfnamefont {M.}~\bibnamefont {Chen}}, \bibinfo
  {author} {\bibfnamefont {S.-M.}\ \bibnamefont {Weng}}, \bibinfo {author}
  {\bibfnamefont {T.-P.}\ \bibnamefont {Yu}}, \bibinfo {author} {\bibfnamefont
  {W.-M.}\ \bibnamefont {Wang}}, \bibinfo {author} {\bibfnamefont
  {F.}~\bibnamefont {He}}, \bibinfo {author} {\bibfnamefont {Z.-M.}\
  \bibnamefont {Sheng}}, \bibinfo {author} {\bibfnamefont {P.}~\bibnamefont
  {McKenna}}, \bibinfo {author} {\bibfnamefont {D.~A.}\ \bibnamefont
  {Jaroszynski}},\ and\ \bibinfo {author} {\bibfnamefont {J.}~\bibnamefont
  {Zhang}},\ }\bibfield  {title} {\bibinfo {title} {Extremely brilliant gev
  $\gamma$-rays from a two-stage laser-plasma accelerator},\ }\href@noop {}
  {\bibfield  {journal} {\bibinfo  {journal} {Science advances}\ }\textbf
  {\bibinfo {volume} {6}},\ \bibinfo {pages} {eaaz7240} (\bibinfo {year}
  {2020})}\BibitemShut {NoStop}%
\bibitem [{\citenamefont {Ruffini}\ \emph {et~al.}(2010)\citenamefont
  {Ruffini}, \citenamefont {Vereshchagin},\ and\ \citenamefont
  {Xue}}]{extreme_environment}%
  \BibitemOpen
  \bibfield  {author} {\bibinfo {author} {\bibfnamefont {R.}~\bibnamefont
  {Ruffini}}, \bibinfo {author} {\bibfnamefont {G.}~\bibnamefont
  {Vereshchagin}},\ and\ \bibinfo {author} {\bibfnamefont {S.-S.}\ \bibnamefont
  {Xue}},\ }\bibfield  {title} {\bibinfo {title} {Electron--positron pairs in
  physics and astrophysics: from heavy nuclei to black holes},\ }\href@noop {}
  {\bibfield  {journal} {\bibinfo  {journal} {Physics Reports}\ }\textbf
  {\bibinfo {volume} {487}},\ \bibinfo {pages} {1} (\bibinfo {year}
  {2010})}\BibitemShut {NoStop}%
\bibitem [{\citenamefont {Bulanov}\ \emph {et~al.}(2015)\citenamefont
  {Bulanov}, \citenamefont {Esirkepov}, \citenamefont {Kando}, \citenamefont
  {Koga}, \citenamefont {Kondo},\ and\ \citenamefont {Korn}}]{RRD_bulanov2015}%
  \BibitemOpen
  \bibfield  {author} {\bibinfo {author} {\bibfnamefont {S.}~\bibnamefont
  {Bulanov}}, \bibinfo {author} {\bibfnamefont {T.~Z.}\ \bibnamefont
  {Esirkepov}}, \bibinfo {author} {\bibfnamefont {M.}~\bibnamefont {Kando}},
  \bibinfo {author} {\bibfnamefont {J.}~\bibnamefont {Koga}}, \bibinfo {author}
  {\bibfnamefont {K.}~\bibnamefont {Kondo}},\ and\ \bibinfo {author}
  {\bibfnamefont {G.}~\bibnamefont {Korn}},\ }\bibfield  {title} {\bibinfo
  {title} {On the problems of relativistic laboratory astrophysics and
  fundamental physics with super powerful lasers},\ }\href@noop {} {\bibfield
  {journal} {\bibinfo  {journal} {Plasma Physics Reports}\ }\textbf {\bibinfo
  {volume} {41}},\ \bibinfo {pages} {1} (\bibinfo {year} {2015})}\BibitemShut
  {NoStop}%
\bibitem [{\citenamefont {Gonzalez-Izquierdo}\ \emph
  {et~al.}(2016)\citenamefont {Gonzalez-Izquierdo}, \citenamefont {Gray},
  \citenamefont {King}, \citenamefont {Dance}, \citenamefont {Wilson},
  \citenamefont {McCreadie}, \citenamefont {Butler}, \citenamefont {Capdessus},
  \citenamefont {Hawkes}, \citenamefont {Green} \emph
  {et~al.}}]{gonzalez2016optically}%
  \BibitemOpen
  \bibfield  {author} {\bibinfo {author} {\bibfnamefont {B.}~\bibnamefont
  {Gonzalez-Izquierdo}}, \bibinfo {author} {\bibfnamefont {R.~J.}\ \bibnamefont
  {Gray}}, \bibinfo {author} {\bibfnamefont {M.}~\bibnamefont {King}}, \bibinfo
  {author} {\bibfnamefont {R.~J.}\ \bibnamefont {Dance}}, \bibinfo {author}
  {\bibfnamefont {R.}~\bibnamefont {Wilson}}, \bibinfo {author} {\bibfnamefont
  {J.}~\bibnamefont {McCreadie}}, \bibinfo {author} {\bibfnamefont {N.~M.}\
  \bibnamefont {Butler}}, \bibinfo {author} {\bibfnamefont {R.}~\bibnamefont
  {Capdessus}}, \bibinfo {author} {\bibfnamefont {S.}~\bibnamefont {Hawkes}},
  \bibinfo {author} {\bibfnamefont {J.~S.}\ \bibnamefont {Green}}, \emph
  {et~al.},\ }\bibfield  {title} {\bibinfo {title} {Optically controlled dense
  current structures driven by relativistic plasma aperture-induced
  diffraction},\ }\href@noop {} {\bibfield  {journal} {\bibinfo  {journal}
  {Nature Physics}\ }\textbf {\bibinfo {volume} {12}},\ \bibinfo {pages} {505}
  (\bibinfo {year} {2016})}\BibitemShut {NoStop}%
\bibitem [{\citenamefont {Di~Piazza}\ \emph {et~al.}(2012)\citenamefont
  {Di~Piazza}, \citenamefont {M{\"u}ller}, \citenamefont {Hatsagortsyan},\ and\
  \citenamefont {Keitel}}]{di2012extremely}%
  \BibitemOpen
  \bibfield  {author} {\bibinfo {author} {\bibfnamefont {A.}~\bibnamefont
  {Di~Piazza}}, \bibinfo {author} {\bibfnamefont {C.}~\bibnamefont
  {M{\"u}ller}}, \bibinfo {author} {\bibfnamefont {K.}~\bibnamefont
  {Hatsagortsyan}},\ and\ \bibinfo {author} {\bibfnamefont {C.~H.}\
  \bibnamefont {Keitel}},\ }\bibfield  {title} {\bibinfo {title} {Extremely
  high-intensity laser interactions with fundamental quantum systems},\
  }\href@noop {} {\bibfield  {journal} {\bibinfo  {journal} {Reviews of Modern
  Physics}\ }\textbf {\bibinfo {volume} {84}},\ \bibinfo {pages} {1177}
  (\bibinfo {year} {2012})}\BibitemShut {NoStop}%
\bibitem [{\citenamefont {Cole}\ \emph {et~al.}(2018)\citenamefont {Cole},
  \citenamefont {Behm}, \citenamefont {Gerstmayr}, \citenamefont {Blackburn},
  \citenamefont {Wood}, \citenamefont {Baird}, \citenamefont {Duff},
  \citenamefont {Harvey}, \citenamefont {Ilderton}, \citenamefont {Joglekar}
  \emph {et~al.}}]{cole2018experimental}%
  \BibitemOpen
  \bibfield  {author} {\bibinfo {author} {\bibfnamefont {J.}~\bibnamefont
  {Cole}}, \bibinfo {author} {\bibfnamefont {K.}~\bibnamefont {Behm}}, \bibinfo
  {author} {\bibfnamefont {E.}~\bibnamefont {Gerstmayr}}, \bibinfo {author}
  {\bibfnamefont {T.}~\bibnamefont {Blackburn}}, \bibinfo {author}
  {\bibfnamefont {J.}~\bibnamefont {Wood}}, \bibinfo {author} {\bibfnamefont
  {C.}~\bibnamefont {Baird}}, \bibinfo {author} {\bibfnamefont {M.~J.}\
  \bibnamefont {Duff}}, \bibinfo {author} {\bibfnamefont {C.}~\bibnamefont
  {Harvey}}, \bibinfo {author} {\bibfnamefont {A.}~\bibnamefont {Ilderton}},
  \bibinfo {author} {\bibfnamefont {A.}~\bibnamefont {Joglekar}}, \emph
  {et~al.},\ }\bibfield  {title} {\bibinfo {title} {Experimental evidence of
  radiation reaction in the collision of a high-intensity laser pulse with a
  laser-wakefield accelerated electron beam},\ }\href@noop {} {\bibfield
  {journal} {\bibinfo  {journal} {Physical Review X}\ }\textbf {\bibinfo
  {volume} {8}},\ \bibinfo {pages} {011020} (\bibinfo {year}
  {2018})}\BibitemShut {NoStop}%
\bibitem [{\citenamefont {Poder}\ \emph {et~al.}(2018)\citenamefont {Poder},
  \citenamefont {Tamburini}, \citenamefont {Sarri}, \citenamefont {Di~Piazza},
  \citenamefont {Kuschel}, \citenamefont {Baird}, \citenamefont {Behm},
  \citenamefont {Bohlen}, \citenamefont {Cole}, \citenamefont {Corvan} \emph
  {et~al.}}]{poder2018experimental}%
  \BibitemOpen
  \bibfield  {author} {\bibinfo {author} {\bibfnamefont {K.}~\bibnamefont
  {Poder}}, \bibinfo {author} {\bibfnamefont {M.}~\bibnamefont {Tamburini}},
  \bibinfo {author} {\bibfnamefont {G.}~\bibnamefont {Sarri}}, \bibinfo
  {author} {\bibfnamefont {A.}~\bibnamefont {Di~Piazza}}, \bibinfo {author}
  {\bibfnamefont {S.}~\bibnamefont {Kuschel}}, \bibinfo {author} {\bibfnamefont
  {C.}~\bibnamefont {Baird}}, \bibinfo {author} {\bibfnamefont
  {K.}~\bibnamefont {Behm}}, \bibinfo {author} {\bibfnamefont {S.}~\bibnamefont
  {Bohlen}}, \bibinfo {author} {\bibfnamefont {J.}~\bibnamefont {Cole}},
  \bibinfo {author} {\bibfnamefont {D.}~\bibnamefont {Corvan}}, \emph
  {et~al.},\ }\bibfield  {title} {\bibinfo {title} {Experimental signatures of
  the quantum nature of radiation reaction in the field of an ultraintense
  laser},\ }\href@noop {} {\bibfield  {journal} {\bibinfo  {journal} {Physical
  Review X}\ }\textbf {\bibinfo {volume} {8}},\ \bibinfo {pages} {031004}
  (\bibinfo {year} {2018})}\BibitemShut {NoStop}%
\bibitem [{\citenamefont {Qu}\ \emph {et~al.}(2021)\citenamefont {Qu},
  \citenamefont {Meuren},\ and\ \citenamefont {Fisch}}]{qu2021signature}%
  \BibitemOpen
  \bibfield  {author} {\bibinfo {author} {\bibfnamefont {K.}~\bibnamefont
  {Qu}}, \bibinfo {author} {\bibfnamefont {S.}~\bibnamefont {Meuren}},\ and\
  \bibinfo {author} {\bibfnamefont {N.~J.}\ \bibnamefont {Fisch}},\ }\bibfield
  {title} {\bibinfo {title} {Signature of collective plasma effects in
  beam-driven qed cascades},\ }\href@noop {} {\bibfield  {journal} {\bibinfo
  {journal} {Physical Review Letters}\ }\textbf {\bibinfo {volume} {127}},\
  \bibinfo {pages} {095001} (\bibinfo {year} {2021})}\BibitemShut {NoStop}%
\bibitem [{\citenamefont {Fedeli}\ \emph {et~al.}(2021)\citenamefont {Fedeli},
  \citenamefont {Sainte-Marie}, \citenamefont {Za{\"\i}m}, \citenamefont
  {Th{\'e}venet}, \citenamefont {Vay}, \citenamefont {Myers}, \citenamefont
  {Qu{\'e}r{\'e}},\ and\ \citenamefont {Vincenti}}]{fedeli2021probing}%
  \BibitemOpen
  \bibfield  {author} {\bibinfo {author} {\bibfnamefont {L.}~\bibnamefont
  {Fedeli}}, \bibinfo {author} {\bibfnamefont {A.}~\bibnamefont
  {Sainte-Marie}}, \bibinfo {author} {\bibfnamefont {N.}~\bibnamefont
  {Za{\"\i}m}}, \bibinfo {author} {\bibfnamefont {M.}~\bibnamefont
  {Th{\'e}venet}}, \bibinfo {author} {\bibfnamefont {J.-L.}\ \bibnamefont
  {Vay}}, \bibinfo {author} {\bibfnamefont {A.}~\bibnamefont {Myers}}, \bibinfo
  {author} {\bibfnamefont {F.}~\bibnamefont {Qu{\'e}r{\'e}}},\ and\ \bibinfo
  {author} {\bibfnamefont {H.}~\bibnamefont {Vincenti}},\ }\bibfield  {title}
  {\bibinfo {title} {Probing strong-field qed with doppler-boosted
  petawatt-class lasers},\ }\href@noop {} {\bibfield  {journal} {\bibinfo
  {journal} {Physical Review Letters}\ }\textbf {\bibinfo {volume} {127}},\
  \bibinfo {pages} {114801} (\bibinfo {year} {2021})}\BibitemShut {NoStop}%
\bibitem [{\citenamefont {Borghesi}\ \emph {et~al.}(2002)\citenamefont
  {Borghesi}, \citenamefont {Campbell}, \citenamefont {Schiavi}, \citenamefont
  {Haines}, \citenamefont {Willi}, \citenamefont {MacKinnon}, \citenamefont
  {Patel}, \citenamefont {Gizzi}, \citenamefont {Galimberti}, \citenamefont
  {Clarke} \emph {et~al.}}]{borghesi2002electric}%
  \BibitemOpen
  \bibfield  {author} {\bibinfo {author} {\bibfnamefont {M.}~\bibnamefont
  {Borghesi}}, \bibinfo {author} {\bibfnamefont {D.}~\bibnamefont {Campbell}},
  \bibinfo {author} {\bibfnamefont {A.}~\bibnamefont {Schiavi}}, \bibinfo
  {author} {\bibfnamefont {M.}~\bibnamefont {Haines}}, \bibinfo {author}
  {\bibfnamefont {O.}~\bibnamefont {Willi}}, \bibinfo {author} {\bibfnamefont
  {A.}~\bibnamefont {MacKinnon}}, \bibinfo {author} {\bibfnamefont
  {P.}~\bibnamefont {Patel}}, \bibinfo {author} {\bibfnamefont
  {L.}~\bibnamefont {Gizzi}}, \bibinfo {author} {\bibfnamefont
  {M.}~\bibnamefont {Galimberti}}, \bibinfo {author} {\bibfnamefont
  {R.}~\bibnamefont {Clarke}}, \emph {et~al.},\ }\bibfield  {title} {\bibinfo
  {title} {Electric field detection in laser-plasma interaction experiments via
  the proton imaging technique},\ }\href@noop {} {\bibfield  {journal}
  {\bibinfo  {journal} {Physics of Plasmas}\ }\textbf {\bibinfo {volume} {9}},\
  \bibinfo {pages} {2214} (\bibinfo {year} {2002})}\BibitemShut {NoStop}%
\bibitem [{\citenamefont {Downer}\ \emph {et~al.}(2018)\citenamefont {Downer},
  \citenamefont {Zgadzaj}, \citenamefont {Debus}, \citenamefont {Schramm},\
  and\ \citenamefont {Kaluza}}]{downer2018diagnostics}%
  \BibitemOpen
  \bibfield  {author} {\bibinfo {author} {\bibfnamefont {M.}~\bibnamefont
  {Downer}}, \bibinfo {author} {\bibfnamefont {R.}~\bibnamefont {Zgadzaj}},
  \bibinfo {author} {\bibfnamefont {A.}~\bibnamefont {Debus}}, \bibinfo
  {author} {\bibfnamefont {U.}~\bibnamefont {Schramm}},\ and\ \bibinfo {author}
  {\bibfnamefont {M.}~\bibnamefont {Kaluza}},\ }\bibfield  {title} {\bibinfo
  {title} {Diagnostics for plasma-based electron accelerators},\ }\href@noop {}
  {\bibfield  {journal} {\bibinfo  {journal} {Reviews of Modern Physics}\
  }\textbf {\bibinfo {volume} {90}},\ \bibinfo {pages} {035002} (\bibinfo
  {year} {2018})}\BibitemShut {NoStop}%
\bibitem [{\citenamefont {Bott}\ \emph {et~al.}(2021)\citenamefont {Bott} \emph
  {et~al.}}]{bott2021PRL}%
  \BibitemOpen
  \bibfield  {author} {\bibinfo {author} {\bibfnamefont {A.~F.}\ \bibnamefont
  {Bott}} \emph {et~al.},\ }\bibfield  {title} {\bibinfo {title} {Inefficient
  magnetic-field amplification in supersonic laser-plasma turbulence},\
  }\href@noop {} {\bibfield  {journal} {\bibinfo  {journal} {Phys. Rev. Lett.}\
  }\textbf {\bibinfo {volume} {127}},\ \bibinfo {pages} {175002} (\bibinfo
  {year} {2021})}\BibitemShut {NoStop}%
\bibitem [{\citenamefont {Gonoskov}\ \emph {et~al.}(2021)\citenamefont
  {Gonoskov}, \citenamefont {Blackburn}, \citenamefont {Marklund},\ and\
  \citenamefont {Bulanov}}]{gonoskov2021charged}%
  \BibitemOpen
  \bibfield  {author} {\bibinfo {author} {\bibfnamefont {A.}~\bibnamefont
  {Gonoskov}}, \bibinfo {author} {\bibfnamefont {T.}~\bibnamefont {Blackburn}},
  \bibinfo {author} {\bibfnamefont {M.}~\bibnamefont {Marklund}},\ and\
  \bibinfo {author} {\bibfnamefont {S.}~\bibnamefont {Bulanov}},\ }\bibfield
  {title} {\bibinfo {title} {Charged particle motion and radiation in strong
  electromagnetic fields},\ }\href@noop {} {\bibfield  {journal} {\bibinfo
  {journal} {arXiv preprint arXiv:2107.02161}\ } (\bibinfo {year}
  {2021})}\BibitemShut {NoStop}%
\bibitem [{\citenamefont {Gonoskov}\ \emph {et~al.}(2014)\citenamefont
  {Gonoskov} \emph {et~al.}}]{gonoskov2014anomalous}%
  \BibitemOpen
  \bibfield  {author} {\bibinfo {author} {\bibfnamefont {A.}~\bibnamefont
  {Gonoskov}} \emph {et~al.},\ }\bibfield  {title} {\bibinfo {title} {Anomalous
  radiative trapping in laser fields of extreme intensity},\ }\href@noop {}
  {\bibfield  {journal} {\bibinfo  {journal} {Physical review letters}\
  }\textbf {\bibinfo {volume} {113}},\ \bibinfo {pages} {014801} (\bibinfo
  {year} {2014})}\BibitemShut {NoStop}%
\bibitem [{\citenamefont {Ji}\ \emph {et~al.}(2014)\citenamefont {Ji},
  \citenamefont {Pukhov}, \citenamefont {Kostyukov}, \citenamefont {Shen},\
  and\ \citenamefont {Akli}}]{ji2014radiation}%
  \BibitemOpen
  \bibfield  {author} {\bibinfo {author} {\bibfnamefont {L.}~\bibnamefont
  {Ji}}, \bibinfo {author} {\bibfnamefont {A.}~\bibnamefont {Pukhov}}, \bibinfo
  {author} {\bibfnamefont {I.~Y.}\ \bibnamefont {Kostyukov}}, \bibinfo {author}
  {\bibfnamefont {B.}~\bibnamefont {Shen}},\ and\ \bibinfo {author}
  {\bibfnamefont {K.}~\bibnamefont {Akli}},\ }\bibfield  {title} {\bibinfo
  {title} {Radiation-reaction trapping of electrons in extreme laser fields},\
  }\href@noop {} {\bibfield  {journal} {\bibinfo  {journal} {Physical review
  letters}\ }\textbf {\bibinfo {volume} {112}},\ \bibinfo {pages} {145003}
  (\bibinfo {year} {2014})}\BibitemShut {NoStop}%
\bibitem [{\citenamefont {Bulanov}\ \emph {et~al.}(2017)\citenamefont {Bulanov}
  \emph {et~al.}}]{bulanov2017charged}%
  \BibitemOpen
  \bibfield  {author} {\bibinfo {author} {\bibfnamefont {S.}~\bibnamefont
  {Bulanov}} \emph {et~al.},\ }\bibfield  {title} {\bibinfo {title} {Charged
  particle dynamics in multiple colliding electromagnetic waves. survey of
  random walk, l{\'e}vy flights, limit circles, attractors and structurally
  determinate patterns},\ }\href@noop {} {\bibfield  {journal} {\bibinfo
  {journal} {Journal of Plasma Physics}\ }\textbf {\bibinfo {volume} {83}}
  (\bibinfo {year} {2017})}\BibitemShut {NoStop}%
\bibitem [{\citenamefont {Bell}\ and\ \citenamefont
  {Kirk}(2008)}]{bell2008possibility}%
  \BibitemOpen
  \bibfield  {author} {\bibinfo {author} {\bibfnamefont {A.}~\bibnamefont
  {Bell}}\ and\ \bibinfo {author} {\bibfnamefont {J.~G.}\ \bibnamefont
  {Kirk}},\ }\bibfield  {title} {\bibinfo {title} {Possibility of prolific pair
  production with high-power lasers},\ }\href@noop {} {\bibfield  {journal}
  {\bibinfo  {journal} {Physical review letters}\ }\textbf {\bibinfo {volume}
  {101}},\ \bibinfo {pages} {200403} (\bibinfo {year} {2008})}\BibitemShut
  {NoStop}%
\bibitem [{\citenamefont {Nerush}\ \emph {et~al.}(2011)\citenamefont {Nerush},
  \citenamefont {Kostyukov}, \citenamefont {Fedotov}, \citenamefont {Narozhny},
  \citenamefont {Elkina},\ and\ \citenamefont {Ruhl}}]{nerush2011laser}%
  \BibitemOpen
  \bibfield  {author} {\bibinfo {author} {\bibfnamefont {E.}~\bibnamefont
  {Nerush}}, \bibinfo {author} {\bibfnamefont {I.~Y.}\ \bibnamefont
  {Kostyukov}}, \bibinfo {author} {\bibfnamefont {A.}~\bibnamefont {Fedotov}},
  \bibinfo {author} {\bibfnamefont {N.}~\bibnamefont {Narozhny}}, \bibinfo
  {author} {\bibfnamefont {N.}~\bibnamefont {Elkina}},\ and\ \bibinfo {author}
  {\bibfnamefont {H.}~\bibnamefont {Ruhl}},\ }\bibfield  {title} {\bibinfo
  {title} {Laser field absorption in self-generated electron-positron pair
  plasma},\ }\href@noop {} {\bibfield  {journal} {\bibinfo  {journal} {Physical
  review letters}\ }\textbf {\bibinfo {volume} {106}},\ \bibinfo {pages}
  {035001} (\bibinfo {year} {2011})}\BibitemShut {NoStop}%
\bibitem [{\citenamefont {Zhu}\ \emph {et~al.}(2016)\citenamefont {Zhu},
  \citenamefont {Yu}, \citenamefont {Sheng}, \citenamefont {Yin}, \citenamefont
  {Turcu},\ and\ \citenamefont {Pukhov}}]{zhu2016dense}%
  \BibitemOpen
  \bibfield  {author} {\bibinfo {author} {\bibfnamefont {X.-L.}\ \bibnamefont
  {Zhu}}, \bibinfo {author} {\bibfnamefont {T.-P.}\ \bibnamefont {Yu}},
  \bibinfo {author} {\bibfnamefont {Z.-M.}\ \bibnamefont {Sheng}}, \bibinfo
  {author} {\bibfnamefont {Y.}~\bibnamefont {Yin}}, \bibinfo {author}
  {\bibfnamefont {I.~C.~E.}\ \bibnamefont {Turcu}},\ and\ \bibinfo {author}
  {\bibfnamefont {A.}~\bibnamefont {Pukhov}},\ }\bibfield  {title} {\bibinfo
  {title} {Dense gev electron--positron pairs generated by lasers in
  near-critical-density plasmas},\ }\href@noop {} {\bibfield  {journal}
  {\bibinfo  {journal} {Nature communications}\ }\textbf {\bibinfo {volume}
  {7}},\ \bibinfo {pages} {1} (\bibinfo {year} {2016})}\BibitemShut {NoStop}%
\bibitem [{\citenamefont {Wang}\ \emph
  {et~al.}(2019{\natexlab{a}})\citenamefont {Wang}, \citenamefont {Toncian},
  \citenamefont {Wei},\ and\ \citenamefont {Arefiev}}]{wang2019structured}%
  \BibitemOpen
  \bibfield  {author} {\bibinfo {author} {\bibfnamefont {T.}~\bibnamefont
  {Wang}}, \bibinfo {author} {\bibfnamefont {T.}~\bibnamefont {Toncian}},
  \bibinfo {author} {\bibfnamefont {M.}~\bibnamefont {Wei}},\ and\ \bibinfo
  {author} {\bibfnamefont {A.}~\bibnamefont {Arefiev}},\ }\bibfield  {title}
  {\bibinfo {title} {Structured targets for detection of megatesla-level
  magnetic fields through faraday rotation of xfel beams},\ }\href@noop {}
  {\bibfield  {journal} {\bibinfo  {journal} {Physics of plasmas}\ }\textbf
  {\bibinfo {volume} {26}},\ \bibinfo {pages} {013105} (\bibinfo {year}
  {2019}{\natexlab{a}})}\BibitemShut {NoStop}%
\bibitem [{\citenamefont {Gong}\ \emph {et~al.}(2021)\citenamefont {Gong},
  \citenamefont {Hatsagortsyan},\ and\ \citenamefont
  {Keitel}}]{gong2021diagnosing}%
  \BibitemOpen
  \bibfield  {author} {\bibinfo {author} {\bibfnamefont {Z.}~\bibnamefont
  {Gong}}, \bibinfo {author} {\bibfnamefont {K.~Z.}\ \bibnamefont
  {Hatsagortsyan}},\ and\ \bibinfo {author} {\bibfnamefont {C.~H.}\
  \bibnamefont {Keitel}},\ }\bibfield  {title} {\bibinfo {title} {Retrieving
  transient magnetic fields of ultrarelativistic laser plasma via ejected
  electron polarization},\ }\href@noop {} {\bibfield  {journal} {\bibinfo
  {journal} {Phys. Rev. Lett.}\ }\textbf {\bibinfo {volume} {127}},\ \bibinfo
  {pages} {165002} (\bibinfo {year} {2021})}\BibitemShut {NoStop}%
\bibitem [{\citenamefont {Akiyama}\ \emph {et~al.}(2021)\citenamefont {Akiyama}
  \emph {et~al.}}]{BH_image_2021_March}%
  \BibitemOpen
  \bibfield  {author} {\bibinfo {author} {\bibfnamefont {K.}~\bibnamefont
  {Akiyama}} \emph {et~al.},\ }\bibfield  {title} {\bibinfo {title} {First m87
  event horizon telescope results. vii. polarization of the ring},\ }\href@noop
  {} {\bibfield  {journal} {\bibinfo  {journal} {The Astrophysical Journal
  Letters}\ }\textbf {\bibinfo {volume} {910}},\ \bibinfo {pages} {L12}
  (\bibinfo {year} {2021})}\BibitemShut {NoStop}%
\bibitem [{\citenamefont {Lembo}\ \emph {et~al.}(2021)\citenamefont {Lembo}
  \emph {et~al.}}]{lembo2021cosmic}%
  \BibitemOpen
  \bibfield  {author} {\bibinfo {author} {\bibfnamefont {M.}~\bibnamefont
  {Lembo}} \emph {et~al.},\ }\bibfield  {title} {\bibinfo {title} {Cosmic
  microwave background polarization as a tool to constrain the optical
  properties of the universe},\ }\href@noop {} {\bibfield  {journal} {\bibinfo
  {journal} {Physical Review Letters}\ }\textbf {\bibinfo {volume} {127}},\
  \bibinfo {pages} {011301} (\bibinfo {year} {2021})}\BibitemShut {NoStop}%
\bibitem [{\citenamefont {Faraday}(1846)}]{faraday1846magnetization}%
  \BibitemOpen
  \bibfield  {author} {\bibinfo {author} {\bibfnamefont {M.}~\bibnamefont
  {Faraday}},\ }\bibfield  {title} {\bibinfo {title} {On the magnetization of
  light and the illumination of magnetic lines of force},\ }\href@noop {}
  {\bibfield  {journal} {\bibinfo  {journal} {Philosophical Transactions of the
  Royal Society of London}\ }\textbf {\bibinfo {volume} {136}},\ \bibinfo
  {pages} {1} (\bibinfo {year} {1846})}\BibitemShut {NoStop}%
\bibitem [{\citenamefont {Chen}(2012)}]{chen2012introduction}%
  \BibitemOpen
  \bibfield  {author} {\bibinfo {author} {\bibfnamefont {F.~F.}\ \bibnamefont
  {Chen}},\ }\href@noop {} {\emph {\bibinfo {title} {Introduction to plasma
  physics}}}\ (\bibinfo  {publisher} {Springer Science \& Business Media},\
  \bibinfo {year} {2012})\BibitemShut {NoStop}%
\bibitem [{\citenamefont {Zhang}\ \emph {et~al.}(2019)\citenamefont {Zhang},
  \citenamefont {Kole}, \citenamefont {Bao}, \citenamefont {Batsch},
  \citenamefont {Bernasconi}, \citenamefont {Cadoux}, \citenamefont {Chai},
  \citenamefont {Dai}, \citenamefont {Dong}, \citenamefont {Gauvin} \emph
  {et~al.}}]{zhang2019detailed}%
  \BibitemOpen
  \bibfield  {author} {\bibinfo {author} {\bibfnamefont {S.-N.}\ \bibnamefont
  {Zhang}}, \bibinfo {author} {\bibfnamefont {M.}~\bibnamefont {Kole}},
  \bibinfo {author} {\bibfnamefont {T.-W.}\ \bibnamefont {Bao}}, \bibinfo
  {author} {\bibfnamefont {T.}~\bibnamefont {Batsch}}, \bibinfo {author}
  {\bibfnamefont {T.}~\bibnamefont {Bernasconi}}, \bibinfo {author}
  {\bibfnamefont {F.}~\bibnamefont {Cadoux}}, \bibinfo {author} {\bibfnamefont
  {J.-Y.}\ \bibnamefont {Chai}}, \bibinfo {author} {\bibfnamefont {Z.-G.}\
  \bibnamefont {Dai}}, \bibinfo {author} {\bibfnamefont {Y.-W.}\ \bibnamefont
  {Dong}}, \bibinfo {author} {\bibfnamefont {N.}~\bibnamefont {Gauvin}}, \emph
  {et~al.},\ }\bibfield  {title} {\bibinfo {title} {Detailed polarization
  measurements of the prompt emission of five gamma-ray bursts},\ }\href@noop
  {} {\bibfield  {journal} {\bibinfo  {journal} {Nature Astronomy}\ }\textbf
  {\bibinfo {volume} {3}},\ \bibinfo {pages} {258} (\bibinfo {year}
  {2019})}\BibitemShut {NoStop}%
\bibitem [{\citenamefont {Gill}\ \emph {et~al.}(2020)\citenamefont {Gill},
  \citenamefont {Granot},\ and\ \citenamefont {Kumar}}]{gill2020linear}%
  \BibitemOpen
  \bibfield  {author} {\bibinfo {author} {\bibfnamefont {R.}~\bibnamefont
  {Gill}}, \bibinfo {author} {\bibfnamefont {J.}~\bibnamefont {Granot}},\ and\
  \bibinfo {author} {\bibfnamefont {P.}~\bibnamefont {Kumar}},\ }\bibfield
  {title} {\bibinfo {title} {Linear polarization in gamma-ray burst prompt
  emission},\ }\href@noop {} {\bibfield  {journal} {\bibinfo  {journal}
  {Monthly Notices of the Royal Astronomical Society}\ }\textbf {\bibinfo
  {volume} {491}},\ \bibinfo {pages} {3343} (\bibinfo {year}
  {2020})}\BibitemShut {NoStop}%
\bibitem [{\citenamefont {Batebi}\ \emph {et~al.}(2016)\citenamefont {Batebi},
  \citenamefont {Mohammadi}, \citenamefont {Ruffini}, \citenamefont
  {Tizchang},\ and\ \citenamefont {Xue}}]{batebi2016generation}%
  \BibitemOpen
  \bibfield  {author} {\bibinfo {author} {\bibfnamefont {S.}~\bibnamefont
  {Batebi}}, \bibinfo {author} {\bibfnamefont {R.}~\bibnamefont {Mohammadi}},
  \bibinfo {author} {\bibfnamefont {R.}~\bibnamefont {Ruffini}}, \bibinfo
  {author} {\bibfnamefont {S.}~\bibnamefont {Tizchang}},\ and\ \bibinfo
  {author} {\bibfnamefont {S.-S.}\ \bibnamefont {Xue}},\ }\bibfield  {title}
  {\bibinfo {title} {Generation of circular polarization of gamma ray bursts},\
  }\href@noop {} {\bibfield  {journal} {\bibinfo  {journal} {Physical Review
  D}\ }\textbf {\bibinfo {volume} {94}},\ \bibinfo {pages} {065033} (\bibinfo
  {year} {2016})}\BibitemShut {NoStop}%
\bibitem [{\citenamefont {B{\oe}hm}\ \emph {et~al.}(2017)\citenamefont
  {B{\oe}hm}, \citenamefont {Degrande}, \citenamefont {Mattelaer},\ and\
  \citenamefont {Vincent}}]{boehm2017circular}%
  \BibitemOpen
  \bibfield  {author} {\bibinfo {author} {\bibfnamefont {C.}~\bibnamefont
  {B{\oe}hm}}, \bibinfo {author} {\bibfnamefont {C.}~\bibnamefont {Degrande}},
  \bibinfo {author} {\bibfnamefont {O.}~\bibnamefont {Mattelaer}},\ and\
  \bibinfo {author} {\bibfnamefont {A.~C.}\ \bibnamefont {Vincent}},\
  }\bibfield  {title} {\bibinfo {title} {Circular polarisation: a new probe of
  dark matter and neutrinos in the sky},\ }\href@noop {} {\bibfield  {journal}
  {\bibinfo  {journal} {Journal of Cosmology and Astroparticle Physics}\
  }\textbf {\bibinfo {volume} {2017}}\bibinfo  {number} { (05)},\ \bibinfo
  {pages} {043}}\BibitemShut {NoStop}%
\bibitem [{\citenamefont {Dean}\ \emph {et~al.}(2008)\citenamefont {Dean},
  \citenamefont {Clark}, \citenamefont {Stephen}, \citenamefont {McBride},
  \citenamefont {Bassani}, \citenamefont {Bazzano}, \citenamefont {Bird},
  \citenamefont {Hill}, \citenamefont {Shaw},\ and\ \citenamefont
  {Ubertini}}]{dean2008polarized}%
  \BibitemOpen
\bibfield  {number} {  }\bibfield  {author} {\bibinfo {author} {\bibfnamefont
  {A.}~\bibnamefont {Dean}}, \bibinfo {author} {\bibfnamefont {D.}~\bibnamefont
  {Clark}}, \bibinfo {author} {\bibfnamefont {J.}~\bibnamefont {Stephen}},
  \bibinfo {author} {\bibfnamefont {V.}~\bibnamefont {McBride}}, \bibinfo
  {author} {\bibfnamefont {L.}~\bibnamefont {Bassani}}, \bibinfo {author}
  {\bibfnamefont {A.}~\bibnamefont {Bazzano}}, \bibinfo {author} {\bibfnamefont
  {A.}~\bibnamefont {Bird}}, \bibinfo {author} {\bibfnamefont {A.}~\bibnamefont
  {Hill}}, \bibinfo {author} {\bibfnamefont {S.}~\bibnamefont {Shaw}},\ and\
  \bibinfo {author} {\bibfnamefont {P.}~\bibnamefont {Ubertini}},\ }\bibfield
  {title} {\bibinfo {title} {Polarized gamma-ray emission from the crab},\
  }\href@noop {} {\bibfield  {journal} {\bibinfo  {journal} {Science}\ }\textbf
  {\bibinfo {volume} {321}},\ \bibinfo {pages} {1183} (\bibinfo {year}
  {2008})}\BibitemShut {NoStop}%
\bibitem [{\citenamefont {Pukhov}(2002)}]{pukhov2002strong}%
  \BibitemOpen
  \bibfield  {author} {\bibinfo {author} {\bibfnamefont {A.}~\bibnamefont
  {Pukhov}},\ }\bibfield  {title} {\bibinfo {title} {Strong field interaction
  of laser radiation},\ }\href@noop {} {\bibfield  {journal} {\bibinfo
  {journal} {Reports on progress in Physics}\ }\textbf {\bibinfo {volume}
  {66}},\ \bibinfo {pages} {47} (\bibinfo {year} {2002})}\BibitemShut {NoStop}%
\bibitem [{\citenamefont {Gong}\ \emph {et~al.}(2020)\citenamefont {Gong},
  \citenamefont {Mackenroth}, \citenamefont {Wang}, \citenamefont {Yan},
  \citenamefont {Toncian},\ and\ \citenamefont {Arefiev}}]{gong2020direct}%
  \BibitemOpen
  \bibfield  {author} {\bibinfo {author} {\bibfnamefont {Z.}~\bibnamefont
  {Gong}}, \bibinfo {author} {\bibfnamefont {F.}~\bibnamefont {Mackenroth}},
  \bibinfo {author} {\bibfnamefont {T.}~\bibnamefont {Wang}}, \bibinfo {author}
  {\bibfnamefont {X.}~\bibnamefont {Yan}}, \bibinfo {author} {\bibfnamefont
  {T.}~\bibnamefont {Toncian}},\ and\ \bibinfo {author} {\bibfnamefont
  {A.}~\bibnamefont {Arefiev}},\ }\bibfield  {title} {\bibinfo {title} {Direct
  laser acceleration of electrons assisted by strong laser-driven azimuthal
  plasma magnetic fields},\ }\href@noop {} {\bibfield  {journal} {\bibinfo
  {journal} {Physical Review E}\ }\textbf {\bibinfo {volume} {102}},\ \bibinfo
  {pages} {013206} (\bibinfo {year} {2020})}\BibitemShut {NoStop}%
\bibitem [{SM()}]{SM}%
  \BibitemOpen
  \href@noop {} {}\bibinfo {howpublished} {See the Supplemental Materials for
  the detailed numerical models, analytical derivation, and extra simulation
  results. The supplemental Materials include
  Refs.~\cite{thomas1927kinematics,bargmann1959precession,duclous2010monte,elkina2011qed,ridgers2014modelling,gonoskov2015extended,li2020_photon,chen2019polarized,li2020production,xue2020generation,arefiev2020energy,robinson2009relativistically,gibbon2005short}.}\BibitemShut
  {Stop}%
\bibitem [{\citenamefont {Arber}\ \emph {et~al.}(2015)\citenamefont {Arber},
  \citenamefont {Bennett}, \citenamefont {Brady}, \citenamefont
  {Lawrence-Douglas}, \citenamefont {Ramsay}, \citenamefont {Sircombe},
  \citenamefont {Gillies}, \citenamefont {Evans}, \citenamefont {Schmitz},
  \citenamefont {Bell} \emph {et~al.}}]{arber2015contemporary}%
  \BibitemOpen
  \bibfield  {author} {\bibinfo {author} {\bibfnamefont {T.}~\bibnamefont
  {Arber}}, \bibinfo {author} {\bibfnamefont {K.}~\bibnamefont {Bennett}},
  \bibinfo {author} {\bibfnamefont {C.}~\bibnamefont {Brady}}, \bibinfo
  {author} {\bibfnamefont {A.}~\bibnamefont {Lawrence-Douglas}}, \bibinfo
  {author} {\bibfnamefont {M.}~\bibnamefont {Ramsay}}, \bibinfo {author}
  {\bibfnamefont {N.}~\bibnamefont {Sircombe}}, \bibinfo {author}
  {\bibfnamefont {P.}~\bibnamefont {Gillies}}, \bibinfo {author} {\bibfnamefont
  {R.}~\bibnamefont {Evans}}, \bibinfo {author} {\bibfnamefont
  {H.}~\bibnamefont {Schmitz}}, \bibinfo {author} {\bibfnamefont
  {A.}~\bibnamefont {Bell}}, \emph {et~al.},\ }\bibfield  {title} {\bibinfo
  {title} {Contemporary particle-in-cell approach to laser-plasma modelling},\
  }\href@noop {} {\bibfield  {journal} {\bibinfo  {journal} {Plasma Physics and
  Controlled Fusion}\ }\textbf {\bibinfo {volume} {57}},\ \bibinfo {pages}
  {113001} (\bibinfo {year} {2015})}\BibitemShut {NoStop}%
\bibitem [{\citenamefont {Liu}\ \emph {et~al.}(2013)\citenamefont {Liu},
  \citenamefont {Wang}, \citenamefont {Liu}, \citenamefont {Fu}, \citenamefont
  {Xu}, \citenamefont {Yan},\ and\ \citenamefont {He}}]{liu2013generating}%
  \BibitemOpen
  \bibfield  {author} {\bibinfo {author} {\bibfnamefont {B.}~\bibnamefont
  {Liu}}, \bibinfo {author} {\bibfnamefont {H.}~\bibnamefont {Wang}}, \bibinfo
  {author} {\bibfnamefont {J.}~\bibnamefont {Liu}}, \bibinfo {author}
  {\bibfnamefont {L.}~\bibnamefont {Fu}}, \bibinfo {author} {\bibfnamefont
  {Y.}~\bibnamefont {Xu}}, \bibinfo {author} {\bibfnamefont {X.}~\bibnamefont
  {Yan}},\ and\ \bibinfo {author} {\bibfnamefont {X.}~\bibnamefont {He}},\
  }\bibfield  {title} {\bibinfo {title} {Generating overcritical dense
  relativistic electron beams via self-matching resonance acceleration},\
  }\href@noop {} {\bibfield  {journal} {\bibinfo  {journal} {Physical review
  letters}\ }\textbf {\bibinfo {volume} {110}},\ \bibinfo {pages} {045002}
  (\bibinfo {year} {2013})}\BibitemShut {NoStop}%
\bibitem [{\citenamefont {Pukhov}\ \emph {et~al.}(1999)\citenamefont {Pukhov},
  \citenamefont {Sheng},\ and\ \citenamefont {Meyer-ter
  Vehn}}]{pukhov1999_DLA}%
  \BibitemOpen
  \bibfield  {author} {\bibinfo {author} {\bibfnamefont {A.}~\bibnamefont
  {Pukhov}}, \bibinfo {author} {\bibfnamefont {Z.-M.}\ \bibnamefont {Sheng}},\
  and\ \bibinfo {author} {\bibfnamefont {J.}~\bibnamefont {Meyer-ter Vehn}},\
  }\bibfield  {title} {\bibinfo {title} {Particle acceleration in relativistic
  laser channels},\ }\href@noop {} {\bibfield  {journal} {\bibinfo  {journal}
  {Physics of Plasmas}\ }\textbf {\bibinfo {volume} {6}},\ \bibinfo {pages}
  {2847} (\bibinfo {year} {1999})}\BibitemShut {NoStop}%
\bibitem [{\citenamefont {Liu}\ \emph {et~al.}(2015)\citenamefont {Liu} \emph
  {et~al.}}]{liu2015quasimonoenergetic}%
  \BibitemOpen
  \bibfield  {author} {\bibinfo {author} {\bibfnamefont {B.}~\bibnamefont
  {Liu}} \emph {et~al.},\ }\bibfield  {title} {\bibinfo {title}
  {Quasimonoenergetic electron beam and brilliant gamma-ray radiation generated
  from near critical density plasma due to relativistic resonant phase
  locking},\ }\href@noop {} {\bibfield  {journal} {\bibinfo  {journal} {Physics
  of Plasmas}\ }\textbf {\bibinfo {volume} {22}},\ \bibinfo {pages} {080704}
  (\bibinfo {year} {2015})}\BibitemShut {NoStop}%
\bibitem [{\citenamefont {Tatischeff}\ \emph {et~al.}(2017)\citenamefont
  {Tatischeff} \emph {et~al.}}]{tatischeff2017astrogam}%
  \BibitemOpen
  \bibfield  {author} {\bibinfo {author} {\bibfnamefont {V.}~\bibnamefont
  {Tatischeff}} \emph {et~al.},\ }\bibfield  {title} {\bibinfo {title}
  {e-astrogam mission: a major step forward for gamma-ray polarimetry},\
  }\href@noop {} {\bibfield  {journal} {\bibinfo  {journal} {Journal of
  Astronomical Telescopes, Instruments, and Systems}\ }\textbf {\bibinfo
  {volume} {4}},\ \bibinfo {pages} {011003} (\bibinfo {year}
  {2017})}\BibitemShut {NoStop}%
\bibitem [{\citenamefont {Rokujo}\ \emph {et~al.}(2018)\citenamefont {Rokujo},
  \citenamefont {Aoki}, \citenamefont {Hamada}, \citenamefont {Hara},
  \citenamefont {Inoue}, \citenamefont {Ishiguro}, \citenamefont {Iyono},
  \citenamefont {Kawahara}, \citenamefont {Kodama}, \citenamefont {Komatani}
  \emph {et~al.}}]{rokujo2018first}%
  \BibitemOpen
  \bibfield  {author} {\bibinfo {author} {\bibfnamefont {H.}~\bibnamefont
  {Rokujo}}, \bibinfo {author} {\bibfnamefont {S.}~\bibnamefont {Aoki}},
  \bibinfo {author} {\bibfnamefont {K.}~\bibnamefont {Hamada}}, \bibinfo
  {author} {\bibfnamefont {T.}~\bibnamefont {Hara}}, \bibinfo {author}
  {\bibfnamefont {T.}~\bibnamefont {Inoue}}, \bibinfo {author} {\bibfnamefont
  {K.}~\bibnamefont {Ishiguro}}, \bibinfo {author} {\bibfnamefont
  {A.}~\bibnamefont {Iyono}}, \bibinfo {author} {\bibfnamefont
  {H.}~\bibnamefont {Kawahara}}, \bibinfo {author} {\bibfnamefont
  {K.}~\bibnamefont {Kodama}}, \bibinfo {author} {\bibfnamefont
  {R.}~\bibnamefont {Komatani}}, \emph {et~al.},\ }\bibfield  {title} {\bibinfo
  {title} {First demonstration of gamma-ray imaging using a balloon-borne
  emulsion telescope},\ }\href@noop {} {\bibfield  {journal} {\bibinfo
  {journal} {Progress of Theoretical and Experimental Physics}\ }\textbf
  {\bibinfo {volume} {2018}},\ \bibinfo {pages} {063H01} (\bibinfo {year}
  {2018})}\BibitemShut {NoStop}%
\bibitem [{\citenamefont {Qiao}\ \emph {et~al.}(2009)\citenamefont {Qiao},
  \citenamefont {Zepf}, \citenamefont {Borghesi},\ and\ \citenamefont
  {Geissler}}]{qiao2009stable}%
  \BibitemOpen
  \bibfield  {author} {\bibinfo {author} {\bibfnamefont {B.}~\bibnamefont
  {Qiao}}, \bibinfo {author} {\bibfnamefont {M.}~\bibnamefont {Zepf}}, \bibinfo
  {author} {\bibfnamefont {M.}~\bibnamefont {Borghesi}},\ and\ \bibinfo
  {author} {\bibfnamefont {M.}~\bibnamefont {Geissler}},\ }\bibfield  {title}
  {\bibinfo {title} {Stable gev ion-beam acceleration from thin foils by
  circularly polarized laser pulses},\ }\href@noop {} {\bibfield  {journal}
  {\bibinfo  {journal} {Physical review letters}\ }\textbf {\bibinfo {volume}
  {102}},\ \bibinfo {pages} {145002} (\bibinfo {year} {2009})}\BibitemShut
  {NoStop}%
\bibitem [{\citenamefont {Chen}\ \emph {et~al.}(2010)\citenamefont {Chen},
  \citenamefont {Pukhov}, \citenamefont {Yu},\ and\ \citenamefont
  {Sheng}}]{chen2010radiation}%
  \BibitemOpen
  \bibfield  {author} {\bibinfo {author} {\bibfnamefont {M.}~\bibnamefont
  {Chen}}, \bibinfo {author} {\bibfnamefont {A.}~\bibnamefont {Pukhov}},
  \bibinfo {author} {\bibfnamefont {T.-P.}\ \bibnamefont {Yu}},\ and\ \bibinfo
  {author} {\bibfnamefont {Z.-M.}\ \bibnamefont {Sheng}},\ }\bibfield  {title}
  {\bibinfo {title} {Radiation reaction effects on ion acceleration in laser
  foil interaction},\ }\href@noop {} {\bibfield  {journal} {\bibinfo  {journal}
  {Plasma Physics and Controlled Fusion}\ }\textbf {\bibinfo {volume} {53}},\
  \bibinfo {pages} {014004} (\bibinfo {year} {2010})}\BibitemShut {NoStop}%
\bibitem [{\citenamefont {Tamburini}\ \emph {et~al.}(2010)\citenamefont
  {Tamburini}, \citenamefont {Pegoraro}, \citenamefont {Di~Piazza},
  \citenamefont {Keitel},\ and\ \citenamefont
  {Macchi}}]{tamburini2010radiation}%
  \BibitemOpen
  \bibfield  {author} {\bibinfo {author} {\bibfnamefont {M.}~\bibnamefont
  {Tamburini}}, \bibinfo {author} {\bibfnamefont {F.}~\bibnamefont {Pegoraro}},
  \bibinfo {author} {\bibfnamefont {A.}~\bibnamefont {Di~Piazza}}, \bibinfo
  {author} {\bibfnamefont {C.~H.}\ \bibnamefont {Keitel}},\ and\ \bibinfo
  {author} {\bibfnamefont {A.}~\bibnamefont {Macchi}},\ }\bibfield  {title}
  {\bibinfo {title} {Radiation reaction effects on radiation pressure
  acceleration},\ }\href@noop {} {\bibfield  {journal} {\bibinfo  {journal}
  {New Journal of Physics}\ }\textbf {\bibinfo {volume} {12}},\ \bibinfo
  {pages} {123005} (\bibinfo {year} {2010})}\BibitemShut {NoStop}%
\bibitem [{\citenamefont {McIlvenny}\ \emph {et~al.}(2021)\citenamefont
  {McIlvenny}, \citenamefont {Doria}, \citenamefont {Romagnani}, \citenamefont
  {Ahmed}, \citenamefont {Booth}, \citenamefont {Ditter}, \citenamefont
  {Ettlinger}, \citenamefont {Hicks}, \citenamefont {Martin}, \citenamefont
  {Scott} \emph {et~al.}}]{mcilvenny2021selective}%
  \BibitemOpen
  \bibfield  {author} {\bibinfo {author} {\bibfnamefont {A.}~\bibnamefont
  {McIlvenny}}, \bibinfo {author} {\bibfnamefont {D.}~\bibnamefont {Doria}},
  \bibinfo {author} {\bibfnamefont {L.}~\bibnamefont {Romagnani}}, \bibinfo
  {author} {\bibfnamefont {H.}~\bibnamefont {Ahmed}}, \bibinfo {author}
  {\bibfnamefont {N.}~\bibnamefont {Booth}}, \bibinfo {author} {\bibfnamefont
  {E.-J.}\ \bibnamefont {Ditter}}, \bibinfo {author} {\bibfnamefont
  {O.}~\bibnamefont {Ettlinger}}, \bibinfo {author} {\bibfnamefont
  {G.}~\bibnamefont {Hicks}}, \bibinfo {author} {\bibfnamefont
  {P.}~\bibnamefont {Martin}}, \bibinfo {author} {\bibfnamefont {G.~G.}\
  \bibnamefont {Scott}}, \emph {et~al.},\ }\bibfield  {title} {\bibinfo {title}
  {Selective ion acceleration by intense radiation pressure},\ }\href@noop {}
  {\bibfield  {journal} {\bibinfo  {journal} {Physical review letters}\
  }\textbf {\bibinfo {volume} {127}},\ \bibinfo {pages} {194801} (\bibinfo
  {year} {2021})}\BibitemShut {NoStop}%
\bibitem [{\citenamefont {Wang}\ \emph
  {et~al.}(2019{\natexlab{b}})\citenamefont {Wang}, \citenamefont {Zepf},\ and\
  \citenamefont {Rykovanov}}]{wang2019intense}%
  \BibitemOpen
  \bibfield  {author} {\bibinfo {author} {\bibfnamefont {J.}~\bibnamefont
  {Wang}}, \bibinfo {author} {\bibfnamefont {M.}~\bibnamefont {Zepf}},\ and\
  \bibinfo {author} {\bibfnamefont {S.}~\bibnamefont {Rykovanov}},\ }\bibfield
  {title} {\bibinfo {title} {Intense attosecond pulses carrying orbital angular
  momentum using laser plasma interactions},\ }\href@noop {} {\bibfield
  {journal} {\bibinfo  {journal} {Nature communications}\ }\textbf {\bibinfo
  {volume} {10}},\ \bibinfo {pages} {1} (\bibinfo {year}
  {2019}{\natexlab{b}})}\BibitemShut {NoStop}%
\bibitem [{\citenamefont {Yi}(2021)}]{yi2021high}%
  \BibitemOpen
  \bibfield  {author} {\bibinfo {author} {\bibfnamefont {L.}~\bibnamefont
  {Yi}},\ }\bibfield  {title} {\bibinfo {title} {High-harmonic generation and
  spin-orbit interaction of light in a relativistic oscillating window},\
  }\href@noop {} {\bibfield  {journal} {\bibinfo  {journal} {Physical Review
  Letters}\ }\textbf {\bibinfo {volume} {126}},\ \bibinfo {pages} {134801}
  (\bibinfo {year} {2021})}\BibitemShut {NoStop}%
\bibitem [{\citenamefont {Yi}\ \emph {et~al.}(2016)\citenamefont {Yi},
  \citenamefont {Pukhov}, \citenamefont {Luu-Thanh},\ and\ \citenamefont
  {Shen}}]{yi2016bright}%
  \BibitemOpen
  \bibfield  {author} {\bibinfo {author} {\bibfnamefont {L.}~\bibnamefont
  {Yi}}, \bibinfo {author} {\bibfnamefont {A.}~\bibnamefont {Pukhov}}, \bibinfo
  {author} {\bibfnamefont {P.}~\bibnamefont {Luu-Thanh}},\ and\ \bibinfo
  {author} {\bibfnamefont {B.}~\bibnamefont {Shen}},\ }\bibfield  {title}
  {\bibinfo {title} {Bright x-ray source from a laser-driven microplasma
  waveguide},\ }\href@noop {} {\bibfield  {journal} {\bibinfo  {journal}
  {Physical review letters}\ }\textbf {\bibinfo {volume} {116}},\ \bibinfo
  {pages} {115001} (\bibinfo {year} {2016})}\BibitemShut {NoStop}%
\bibitem [{\citenamefont {Wang}\ \emph {et~al.}(2018)\citenamefont {Wang},
  \citenamefont {Sheng}, \citenamefont {Gibbon}, \citenamefont {Chen},
  \citenamefont {Li},\ and\ \citenamefont {Zhang}}]{wang2018collimated}%
  \BibitemOpen
  \bibfield  {author} {\bibinfo {author} {\bibfnamefont {W.-M.}\ \bibnamefont
  {Wang}}, \bibinfo {author} {\bibfnamefont {Z.-M.}\ \bibnamefont {Sheng}},
  \bibinfo {author} {\bibfnamefont {P.}~\bibnamefont {Gibbon}}, \bibinfo
  {author} {\bibfnamefont {L.-M.}\ \bibnamefont {Chen}}, \bibinfo {author}
  {\bibfnamefont {Y.-T.}\ \bibnamefont {Li}},\ and\ \bibinfo {author}
  {\bibfnamefont {J.}~\bibnamefont {Zhang}},\ }\bibfield  {title} {\bibinfo
  {title} {Collimated ultrabright gamma rays from electron wiggling along a
  petawatt laser-irradiated wire in the qed regime},\ }\href@noop {} {\bibfield
   {journal} {\bibinfo  {journal} {Proceedings of the National Academy of
  Sciences}\ }\textbf {\bibinfo {volume} {115}},\ \bibinfo {pages} {9911}
  (\bibinfo {year} {2018})}\BibitemShut {NoStop}%
\bibitem [{\citenamefont {Kaymak}\ \emph {et~al.}(2016)\citenamefont {Kaymak},
  \citenamefont {Pukhov}, \citenamefont {Shlyaptsev},\ and\ \citenamefont
  {Rocca}}]{kaymak2016nanoscale}%
  \BibitemOpen
  \bibfield  {author} {\bibinfo {author} {\bibfnamefont {V.}~\bibnamefont
  {Kaymak}}, \bibinfo {author} {\bibfnamefont {A.}~\bibnamefont {Pukhov}},
  \bibinfo {author} {\bibfnamefont {V.~N.}\ \bibnamefont {Shlyaptsev}},\ and\
  \bibinfo {author} {\bibfnamefont {J.~J.}\ \bibnamefont {Rocca}},\ }\bibfield
  {title} {\bibinfo {title} {Nanoscale ultradense z-pinch formation from
  laser-irradiated nanowire arrays},\ }\href@noop {} {\bibfield  {journal}
  {\bibinfo  {journal} {Physical review letters}\ }\textbf {\bibinfo {volume}
  {117}},\ \bibinfo {pages} {035004} (\bibinfo {year} {2016})}\BibitemShut
  {NoStop}%
\bibitem [{\citenamefont {Grismayer}\ \emph {et~al.}(2016)\citenamefont
  {Grismayer} \emph {et~al.}}]{grismayer2016laser}%
  \BibitemOpen
  \bibfield  {author} {\bibinfo {author} {\bibfnamefont {T.}~\bibnamefont
  {Grismayer}} \emph {et~al.},\ }\bibfield  {title} {\bibinfo {title} {Laser
  absorption via quantum electrodynamics cascades in counter propagating laser
  pulses},\ }\href@noop {} {\bibfield  {journal} {\bibinfo  {journal} {Physics
  of Plasmas}\ }\textbf {\bibinfo {volume} {23}},\ \bibinfo {pages} {056706}
  (\bibinfo {year} {2016})}\BibitemShut {NoStop}%
\bibitem [{\citenamefont {Thomas}(1927)}]{thomas1927kinematics}%
  \BibitemOpen
  \bibfield  {author} {\bibinfo {author} {\bibfnamefont {L.~H.}\ \bibnamefont
  {Thomas}},\ }\bibfield  {title} {\bibinfo {title} {I. the kinematics of an
  electron with an axis},\ }\href@noop {} {\bibfield  {journal} {\bibinfo
  {journal} {The London, Edinburgh, and Dublin Philosophical Magazine and
  Journal of Science}\ }\textbf {\bibinfo {volume} {3}},\ \bibinfo {pages} {1}
  (\bibinfo {year} {1927})}\BibitemShut {NoStop}%
\bibitem [{\citenamefont {Bargmann}\ \emph {et~al.}(1959)\citenamefont
  {Bargmann}, \citenamefont {Michel},\ and\ \citenamefont
  {Telegdi}}]{bargmann1959precession}%
  \BibitemOpen
  \bibfield  {author} {\bibinfo {author} {\bibfnamefont {V.}~\bibnamefont
  {Bargmann}}, \bibinfo {author} {\bibfnamefont {L.}~\bibnamefont {Michel}},\
  and\ \bibinfo {author} {\bibfnamefont {V.}~\bibnamefont {Telegdi}},\
  }\bibfield  {title} {\bibinfo {title} {Precession of the polarization of
  particles moving in a homogeneous electromagnetic field},\ }\href@noop {}
  {\bibfield  {journal} {\bibinfo  {journal} {Physical Review Letters}\
  }\textbf {\bibinfo {volume} {2}},\ \bibinfo {pages} {435} (\bibinfo {year}
  {1959})}\BibitemShut {NoStop}%
\bibitem [{\citenamefont {Duclous}\ \emph {et~al.}(2010)\citenamefont {Duclous}
  \emph {et~al.}}]{duclous2010monte}%
  \BibitemOpen
  \bibfield  {author} {\bibinfo {author} {\bibfnamefont {R.}~\bibnamefont
  {Duclous}} \emph {et~al.},\ }\bibfield  {title} {\bibinfo {title} {Monte
  carlo calculations of pair production in high-intensity laser--plasma
  interactions},\ }\href@noop {} {\bibfield  {journal} {\bibinfo  {journal}
  {Plasma Physics and Controlled Fusion}\ }\textbf {\bibinfo {volume} {53}},\
  \bibinfo {pages} {015009} (\bibinfo {year} {2010})}\BibitemShut {NoStop}%
\bibitem [{\citenamefont {Elkina}\ \emph {et~al.}(2011)\citenamefont {Elkina}
  \emph {et~al.}}]{elkina2011qed}%
  \BibitemOpen
  \bibfield  {author} {\bibinfo {author} {\bibfnamefont {N.}~\bibnamefont
  {Elkina}} \emph {et~al.},\ }\bibfield  {title} {\bibinfo {title} {Qed
  cascades induced by circularly polarized laser fields},\ }\href@noop {}
  {\bibfield  {journal} {\bibinfo  {journal} {Physical Review Special
  Topics-Accelerators and Beams}\ }\textbf {\bibinfo {volume} {14}},\ \bibinfo
  {pages} {054401} (\bibinfo {year} {2011})}\BibitemShut {NoStop}%
\bibitem [{\citenamefont {Ridgers}\ \emph {et~al.}(2014)\citenamefont {Ridgers}
  \emph {et~al.}}]{ridgers2014modelling}%
  \BibitemOpen
  \bibfield  {author} {\bibinfo {author} {\bibfnamefont {C.}~\bibnamefont
  {Ridgers}} \emph {et~al.},\ }\bibfield  {title} {\bibinfo {title} {Modelling
  gamma-ray photon emission and pair production in high-intensity laser--matter
  interactions},\ }\href@noop {} {\bibfield  {journal} {\bibinfo  {journal}
  {Journal of Computational Physics}\ }\textbf {\bibinfo {volume} {260}},\
  \bibinfo {pages} {273} (\bibinfo {year} {2014})}\BibitemShut {NoStop}%
\bibitem [{\citenamefont {Gonoskov}\ \emph {et~al.}(2015)\citenamefont
  {Gonoskov} \emph {et~al.}}]{gonoskov2015extended}%
  \BibitemOpen
  \bibfield  {author} {\bibinfo {author} {\bibfnamefont {A.}~\bibnamefont
  {Gonoskov}} \emph {et~al.},\ }\bibfield  {title} {\bibinfo {title} {Extended
  particle-in-cell schemes for physics in ultrastrong laser fields: Review and
  developments},\ }\href@noop {} {\bibfield  {journal} {\bibinfo  {journal}
  {Physical Review E}\ }\textbf {\bibinfo {volume} {92}},\ \bibinfo {pages}
  {023305} (\bibinfo {year} {2015})}\BibitemShut {NoStop}%
\bibitem [{\citenamefont {Li}\ \emph {et~al.}(2020{\natexlab{a}})\citenamefont
  {Li} \emph {et~al.}}]{li2020_photon}%
  \BibitemOpen
  \bibfield  {author} {\bibinfo {author} {\bibfnamefont {Y.-F.}\ \bibnamefont
  {Li}} \emph {et~al.},\ }\bibfield  {title} {\bibinfo {title} {Polarized
  ultrashort brilliant multi-gev $\gamma$ rays via single-shot laser-electron
  interaction},\ }\href@noop {} {\bibfield  {journal} {\bibinfo  {journal}
  {Physical review letters}\ }\textbf {\bibinfo {volume} {124}},\ \bibinfo
  {pages} {014801} (\bibinfo {year} {2020}{\natexlab{a}})}\BibitemShut
  {NoStop}%
\bibitem [{\citenamefont {Chen}\ \emph {et~al.}(2019)\citenamefont {Chen},
  \citenamefont {He}, \citenamefont {Shaisultanov}, \citenamefont
  {Hatsagortsyan},\ and\ \citenamefont {Keitel}}]{chen2019polarized}%
  \BibitemOpen
  \bibfield  {author} {\bibinfo {author} {\bibfnamefont {Y.-Y.}\ \bibnamefont
  {Chen}}, \bibinfo {author} {\bibfnamefont {P.-L.}\ \bibnamefont {He}},
  \bibinfo {author} {\bibfnamefont {R.}~\bibnamefont {Shaisultanov}}, \bibinfo
  {author} {\bibfnamefont {K.~Z.}\ \bibnamefont {Hatsagortsyan}},\ and\
  \bibinfo {author} {\bibfnamefont {C.~H.}\ \bibnamefont {Keitel}},\ }\bibfield
   {title} {\bibinfo {title} {Polarized positron beams via intense two-color
  laser pulses},\ }\href@noop {} {\bibfield  {journal} {\bibinfo  {journal}
  {Physical review letters}\ }\textbf {\bibinfo {volume} {123}},\ \bibinfo
  {pages} {174801} (\bibinfo {year} {2019})}\BibitemShut {NoStop}%
\bibitem [{\citenamefont {Li}\ \emph {et~al.}(2020{\natexlab{b}})\citenamefont
  {Li} \emph {et~al.}}]{li2020production}%
  \BibitemOpen
  \bibfield  {author} {\bibinfo {author} {\bibfnamefont {Y.-F.}\ \bibnamefont
  {Li}} \emph {et~al.},\ }\bibfield  {title} {\bibinfo {title} {Production of
  highly polarized positron beams via helicity transfer from polarized
  electrons in a strong laser field},\ }\href@noop {} {\bibfield  {journal}
  {\bibinfo  {journal} {Physical Review Letters}\ }\textbf {\bibinfo {volume}
  {125}},\ \bibinfo {pages} {044802} (\bibinfo {year}
  {2020}{\natexlab{b}})}\BibitemShut {NoStop}%
\bibitem [{\citenamefont {Xue}\ \emph {et~al.}(2020)\citenamefont {Xue} \emph
  {et~al.}}]{xue2020generation}%
  \BibitemOpen
  \bibfield  {author} {\bibinfo {author} {\bibfnamefont {K.}~\bibnamefont
  {Xue}} \emph {et~al.},\ }\bibfield  {title} {\bibinfo {title} {Generation of
  highly-polarized high-energy brilliant $\gamma$-rays via laser-plasma
  interaction},\ }\href@noop {} {\bibfield  {journal} {\bibinfo  {journal}
  {Matter and Radiation at Extremes}\ }\textbf {\bibinfo {volume} {5}},\
  \bibinfo {pages} {054402} (\bibinfo {year} {2020})}\BibitemShut {NoStop}%
\bibitem [{\citenamefont {Arefiev}\ \emph {et~al.}(2020)\citenamefont
  {Arefiev}, \citenamefont {Gong},\ and\ \citenamefont
  {Robinson}}]{arefiev2020energy}%
  \BibitemOpen
  \bibfield  {author} {\bibinfo {author} {\bibfnamefont {A.}~\bibnamefont
  {Arefiev}}, \bibinfo {author} {\bibfnamefont {Z.}~\bibnamefont {Gong}},\ and\
  \bibinfo {author} {\bibfnamefont {A.}~\bibnamefont {Robinson}},\ }\bibfield
  {title} {\bibinfo {title} {Energy gain by laser-accelerated electrons in a
  strong magnetic field},\ }\href@noop {} {\bibfield  {journal} {\bibinfo
  {journal} {Physical Review E}\ }\textbf {\bibinfo {volume} {101}},\ \bibinfo
  {pages} {043201} (\bibinfo {year} {2020})}\BibitemShut {NoStop}%
\bibitem [{\citenamefont {Robinson}\ \emph {et~al.}(2009)\citenamefont
  {Robinson}, \citenamefont {Gibbon}, \citenamefont {Zepf}, \citenamefont
  {Kar}, \citenamefont {Evans},\ and\ \citenamefont
  {Bellei}}]{robinson2009relativistically}%
  \BibitemOpen
  \bibfield  {author} {\bibinfo {author} {\bibfnamefont {A.}~\bibnamefont
  {Robinson}}, \bibinfo {author} {\bibfnamefont {P.}~\bibnamefont {Gibbon}},
  \bibinfo {author} {\bibfnamefont {M.}~\bibnamefont {Zepf}}, \bibinfo {author}
  {\bibfnamefont {S.}~\bibnamefont {Kar}}, \bibinfo {author} {\bibfnamefont
  {R.}~\bibnamefont {Evans}},\ and\ \bibinfo {author} {\bibfnamefont
  {C.}~\bibnamefont {Bellei}},\ }\bibfield  {title} {\bibinfo {title}
  {Relativistically correct hole-boring and ion acceleration by circularly
  polarized laser pulses},\ }\href@noop {} {\bibfield  {journal} {\bibinfo
  {journal} {Plasma Physics and Controlled Fusion}\ }\textbf {\bibinfo {volume}
  {51}},\ \bibinfo {pages} {024004} (\bibinfo {year} {2009})}\BibitemShut
  {NoStop}%
\bibitem [{\citenamefont {Gibbon}(2005)}]{gibbon2005short}%
  \BibitemOpen
  \bibfield  {author} {\bibinfo {author} {\bibfnamefont {P.}~\bibnamefont
  {Gibbon}},\ }\href@noop {} {\emph {\bibinfo {title} {Short pulse laser
  interactions with matter}}}\ (\bibinfo  {publisher} {World Scientific},\
  \bibinfo {year} {2005})\BibitemShut {NoStop}%
\end{thebibliography}
%

\end{document}